\documentstyle[11pt,aaspp4]{article}
\tightenlines
\def	\beq	{\begin{equation}}
\def	\bB	{{\bf B}}
\def	\bE	{{\bf E}}
\def	\bH	{{\bf H}}
\def	\bJ	{{\bf J}}
\def	\bM	{{\bf M}}
\def	\cm	{\,{\rm cm}}
\def	\debye	{\,{\rm debye}}
\def	\eeq	{\end{equation}}
\def	\erg	{\,{\rm ergs}}

\def	\g	{\,{\rm g}}
\def	\G	{\,{\rm G}}
\def	\GHz	{\,{\rm GHz}}
\def	\gtsim	{\gtrsim}					  %apj version
\def	\H	{{\rm H}}

\def	\Hz	{\,{\rm Hz}}
\def	\K	{\,{\rm K}}

\def	\ltsim	{\lesssim}					  %apj version
\def	\micron	{\,\mu{\rm m}}
\def	\mm	{\,{\rm mm}}

\def	\s	{\,{\rm s}}

\lefthead{Draine \& Lazarian}
\righthead{Magnetic Dipole Emission from Dust}

\begin{document}

%\begin{center}
%{\bf *** DRAFT \today\ *** Please do not circulate.}
\begin{center}
{\bf POPe-770}
\quad\quad\quad\quad\quad\quad\quad\quad\quad\quad 
Submitted to {\it The Astrophysical Journal} 98.06.29
\end{center}
%\end{center}
\title{Magnetic Dipole Microwave Emission from Dust Grains}

\author{B.T. Draine \& A. Lazarian}
\affil{Princeton University Observatory, Peyton Hall, Princeton,
NJ 08544;\\
draine@astro.princeton.edu, lazarian@astro.princeton.edu}

\begin{abstract}

Thermal fluctuations in the magnetization of interstellar
grains will produce magnetic dipole emission at $\nu\ltsim100\GHz$.
We show how to calculate absorption and emission from
small particles composed of material
with magnetic as well as dielectric properties.
The Kramers-Kronig relations for a dusty medium are generalized to
include the possibility of magnetic grains.

The magnetic permeability as a function of frequency is discussed for several
candidate grain materials.
Iron grains, or grains containing iron inclusions, are likely to
have the magnetic analogue of a Fr\"ohlich resonance in the 
vicinity of  $\sim50-100\GHz$,
resulting in a large magnetic dipole absorption cross section.

We calculate the emission spectra for various interstellar 
grain candidates.
While ``ordinary'' paramagnetic grains or even magnetite grains
cannot account for
the observed ``anomalous'' emission from dust in the $14 - 90 \GHz$ range,
stronger magnetic dipole emission will result if a fraction of the
grain material is ferromagnetic, as could be the case given the
high Fe content of interstellar dust.
The observed emission from dust near $90\GHz$ implies that not more than
$\sim5\%$ of interstellar Fe is in the form of metallic 
iron grains or inclusions (e.g., in ``GEMS'').
However, we show that if most interstellar Fe is in a moderately
ferromagnetic material, with the magnetic properties properly adjusted,
it could contribute a substantial fraction of the observed
$14 - 90 \GHz$ emission, perhaps comparable to the
contribution from spinning ultrasmall dust grains.
The two emission mechanisms can be distinguished by measuring the
emission from dark clouds.

If ferromagnetic grains consist of a single magnetic domain, and are
aligned,
the magnetic dipole emission will be linearly polarized, with
the polarization depending strongly on frequency.

\end{abstract}

\keywords{ISM: Atomic Processes, Dust, Radiation; Cosmic Microwave Background}

\section{Introduction}

Experiments to map the cosmic background radiation (CBR) 
have stimulated renewed
interest in diffuse Galactic emission.
Sensitive observations of variations in the microwave sky brightness have
revealed 14-90 GHz microwave emission
which is correlated with
$100\micron$ thermal emission from interstellar dust
(Kogut et al.\ 1996;
de Oliveira-Costa et al.\ 1997;
Leitch et al.\ 1997).
The origin of this ``anomalous" emission has been of great interest.
While the observed frequency-dependence appears consistent
with free-free emission
(Kogut et al.\ 1996; Leitch et al.\ 1997),
the observed intensities cannot be due to free-free
emission from interstellar gas: 
for gas temperature $T\approx10^4\K$
the required spatial variations in 
emission measure would produce $\sim60$ times more H$\alpha$ 
than allowed by
observations (Gaustad et al. 1996; Leitch et al.\ 1997), 
while $T\approx10^6\K$ gas would produce
powerful X-ray emission which can be ruled out on energetic grounds
(Draine \& Lazarian 1998).

It has been proposed that the observed 14-90 GHz emission is
rotational emission from the population
of ultrasmall dust grains believed to be present in the interstellar
medium (Draine \& Lazarian 1998a,b).
Such electric dipole emission is a natural consequence of current
models of interstellar dust, and seems likely to be the mechanism
responsible for the observed emission.
It is nevertheless of interest to investigate whether other emission
mechanisms exist which would also be capable of producing strong
emission at these frequencies.

Here we discuss an entirely different emission mechanism: 
{\it thermal fluctuations
of the magnetization} within individual interstellar grains.
Such thermal fluctuations
will result in magnetic dipole radiation.
The magnitude of the magnetic fluctuations depends upon the magnetic
properties of the grain material, and we consider a number of possibilities,
including ``ordinary'' paramagnetism as well as 
superparamagnetism, ferromagnetism, and ferrimagnetism.

If about 30\% of the grain mass is carbonaceous, then Fe and Ni contribute
nearly 30\% of the mass of the remaining grain material.
It would not be surprising, then, if some fraction of the interstellar
grain population were strongly magnetic; metallic iron/nickel, magnetite
(Fe$_3$O$_4$), and maghemite ($\gamma$Fe$_2$O$_3$) 
are plausible candidates to consider.

At infrared and optical frequencies ($\nu\gtsim 10^{12}\Hz$)
it has been customary to neglect the
magnetic properties of the grain material when computing absorption or
emission of electromagnetic radiation by dust grains; this is an
excellent approximation since most materials have negligible
magnetic response to oscillating magnetic fields at frequencies
$\gtsim 10^{11}\Hz$.
This is because ``magnetism'' is due to ordering
of electron spins, and the maximum frequency for electron
spins to reorient is the precession frequency of an electron in the
local magnetic field (due mainly to other electron spins) in the
material; these gyrofrequencies do not exceed $\sim 20\GHz$.
In contrast, the electron 
charge distribution in grains can respond strongly to
applied electric fields at frequencies as large as $10^{16}\Hz$.

At microwave frequencies materials can respond to both electric and
magnetic fields,
and
both the dielectric constant $\epsilon(\omega)$ and the magnetic permeability 
$\mu(\omega)$ can play a role in the absorption and emission of electromagnetic
radiation.

The present study is primarily directed at calculation of thermal emission
from magnetic dust grains.
The complex permeability $\mu(\omega)$ obtained here is 
also relevant to the process of
grain alignment via magnetic dissipation (Davis \& Greenstein 1951),
particularly for ultrasmall grains undergoing very rapid rotation
(Draine \& Lazarian 1998a,b; Lazarian \& Draine 1998)
and suprathermally rotating graints (Purcell 1979).

The optical properties of small
particles, including both dielectric and magnetic effects,
are reviewd in \S\ref{sec:optics}.
In \S\ref{sec:KK} we generalize Purcell's (1969) application of the
Kramers-Kronig relations to interstellar grains to include the
case of magnetic grains.

The expected form of the frequency-dependence of the complex permeability
$\mu(\omega)$ is discussed in \S\ref{sec:magnetic}.
In \S\ref{sec:paramag} we review the properties of ordinary paramagnetic
materials, and in
\S\ref{sec:ferro} we discuss various classes of iron-rich materials,
including: single-domain iron grains (\S\ref{sec:Fe_single});
bulk (multidomain) iron (\S\ref{sec:Fe_bulk});
materials with single-domain Fe inclusions (\S\ref{sec:Fe_incl}),
including the superparamagnetic limit where the inclusions contain less
than $\sim10^3$ Fe atoms each (\S\ref{sec:superpara});
ferrimagnetic materials such as magnetite and
maghemite (\S\ref{sec:ferri});
and antiferromagnetic materials (\S\ref{sec:antiferro}).

In \S\ref{sec:prediction} 
we consider several ways in which iron can be present in the interstellar
grain population, and for each case we calculate the expected thermal
emission contributed by magnetic fluctuations.
Iron grains, if present, would produce strong $50-100\GHz$ emission; this
can be used to place an upper limit of $\sim5\%$ on the fraction of the
interstellar Fe which can be in the form of pure iron grains or inclusions.
In \S\ref{sec:hypothetical_materials} 
we discuss the possibility that the Fe could be in a less strongly magnetic
form (e.g., impure iron), and show that there are plausible magnetic
properties for which such grains could dominate the 14-90 GHz emission.
The polarization of the magnetic dipole emission from ferromagnetic grains is
discussed in \S\ref{sec:polarization}.

We discuss our results in \S\ref{sec:discussion}, and
summarize in \S\ref{sec:summary}.

\section{Optics of Magnetic Grains\label{sec:optics}}
\subsection{Spherical Grains\label{sec:spheres}}
For grains which are small compared to the wavelength, 
the cross section for absorption of electromagnetic waves 
can be written as the sum of electric dipole
and magnetic dipole
cross sections
(Draine \& Lee 1984):
\beq
C_{abs} \approx C_{abs}^{(ed)} + C_{abs}^{(md)} ~~~.
\label{eq:dipoleapprox}
\eeq
For nonmagnetic materials (i.e., materials with magnetic permeability 
$\mu=1$), 
the magnetic dipole contribution arises from
induced eddy currents in the grain, which give it an oscillating
magnetic dipole moment which is out of phase with the applied
magnetic field; the absorption cross section
for this case has been given previously (Landau \& Lifshitz 1960;
Draine \& Lee 1984).
For materials with a magnetic as well as electric response, the
absorption cross section for small spheres of radius $a$ can be shown to be
\beq
C_{abs}^{(ed)}=
V {\omega \over c}
\left\{
{
9\epsilon_2 \over 
|\epsilon+2|^2
}
+
\left({\omega a\over c}\right)^2
\left|{3\epsilon\over\epsilon+2}\right|^2
{\mu_2\over 10}
\right\} ~~~,
\label{eq:cabs_ed}
\eeq
\beq
C_{abs}^{(md)}=
V {\omega \over c}
\left\{
{
9\mu_2 \over 
|\mu+2|^2
}
+ 
\left(\omega a \over c\right)^2
\left|{3\mu\over\mu+2}\right|^2
{\epsilon_2\over 10}
\right\} ~~~,
\label{eq:cabs_md}
\eeq
where $V=4\pi a^3/3$, $\epsilon\equiv\epsilon_1+i\epsilon_2$ is 
the dielectric constant,\footnote{
	The electrical conductivity $\sigma=\omega\epsilon_2/4\pi$.
	}
and $\mu\equiv 1+4\pi\chi\equiv\mu_1+i\mu_2$ is the magnetic permeability.
The dipole approximation (\ref{eq:dipoleapprox}) is valid\footnote{
	Eq.\ (\protect{\ref{eq:cabs_ed}}) neglects ``shielding'' of the
	grain interior from the applied magnetic field by induced
	eddy currents, and (\protect{\ref{eq:cabs_md}}) neglects
	shielding from the applied electric field by induced
	time-dependent magnetization.
	It can be shown that such shielding is negligible provided
	the dipole validity criterion (\protect{\ref{eq:dipolevalidity}})
	is satisfied.
	}
provided
\beq
|\epsilon\mu|^{1/2}
\left({\omega a\over c}\right) < 1
\label{eq:dipolevalidity} ~~~.
\eeq
In eq. (\ref{eq:cabs_ed}) the first term is due to the polarization in 
response to the
applied electric field, and the second term is due to the
oscillating circular magnetization induced by the time-dependent
displacement current $\partial {\bf D}/\partial t$.
Similarly, in eq. (\ref{eq:cabs_md}) the first term is due to the
magnetization in response to the applied magnetic field, while
the second term is the heating due to the
``eddy currents'' induced by the time-dependent magnetic field
$\partial {\bf B}/\partial t$.
By analogy, we will refer to the second term in eq.\ (\ref{eq:cabs_ed})
as being due to ``eddy magnetization''.
Under some circumstances the second-order ``eddy''
contributions can be important -- for example, 
``eddy current'' dissipation is important for $a\gtsim0.2\micron$
graphite grains at $\sim3\times10^{13}\Hz$ (Draine \& Lee 1984).
However,
for $a\ltsim 3\times10^{-5}\cm$ interstellar grains at
$\nu\ltsim10^{12}\Hz$, 
$(\omega a/c)^2 < 4\times10^{-5}$ and these higher-order 
terms are generally negligible.
The ``eddy'' terms will therefore be neglected in the remainder of this
paper.
Thus,
\beq
{C_{abs}^{(md)} \over C_{abs}^{(ed)}} \approx
{\mu_2 |\epsilon+2|^2\over\epsilon_2 |\mu+2|^2}
\label{eq:cabsmd_over_cabsed}~~~.
\eeq

\subsection{Ellipsoidal Grains\label{sec:ellipsoids}}

Ellipsoidal grains with semiaxes $a_1\leq a_2 \leq a_3$ 
are characterized by 
``geometrical factors'' or ``depolarization factors''
$L_1\geq L_2\geq L_3$ which depend only on the grain shape
(see Bohren \& Huffman 1983).
Table \ref{tab:geom_fact} contains $L_j$ for selected
ellipsoidal shapes.

%\placetable{tab:geom_fact}
%{\begin{center}[Editor: Place Table \ref{tab:geom_fact} here.]\end{center}}
\begin{table}
\begin{center}

\begin{tabular}{|c|c|c|c|c|}
\hline
$a_1:a_2:a_3$	& $L_1$	& $L_2$ & $L_3$	&Note\\
\hline
$1:1:1$		&0.3333333	&0.3333333	&0.3333333	&sphere\\
$1:1:1.25$	&0.3620042	&0.3620042	&0.2759916 	&prolate spheroid\\
$1:1:1.5$	&0.3835093	&0.3835093	&0.2329815	&prolate spheroid\\
$1:1:2$		&0.4132180	&0.4132180	&0.1735640	&prolate spheroid\\
$1:1.25:1.25$	&0.3944403	&0.3027798	&0.3027798 	&oblate spheroid\\
$1:1.25:1.5$	&0.4189528	&0.3233250	&0.2577222 	&\\
$1:1.5:1.5$	&0.4459056	&0.2770472	&0.2770472 	&oblate spheroid\\
$1:1.5:2$	&0.4837282	&0.3050063	&0.2112656	&\\
$1:2:2$		&0.5272003	&0.2363999	&0.2363999	&oblate spheroid\\
\hline
\end{tabular}
\end{center}
\caption{\label{tab:geom_fact}
	Geometrical factors $L_j$ for ellipsoids with semiaxes $a_j$.
	}
\end{table}

For the moment, we assume $\epsilon$ and $\mu$ to be isotropic;
we will discuss anisotropic $\mu$ in \S\ref{sec:polarization} below.
The absorption cross section for incident radiation with the
electric vector parallel to semiaxis $e$, and magnetic vector parallel
to semiaxis $h$, is
\beq
C_{abs}^{(ed)} \approx V {\omega \over c}
\left[
{\epsilon_2 \over |1+ L_e(\epsilon-1)|^2}
\right] ~~~,
\label{eq:cabs_ed_ell}
\eeq
\beq
C_{abs}^{(md)} \approx V {\omega \over c}
\left[
{\mu_2 \over |1+L_h(\mu-1)|^2}
\right] ~~~,
\label{eq:cabs_md_ell}
\eeq
where $V\equiv(4\pi/3) a_1 a_2 a_3$, and
we have neglected the ``eddy current'' and ``eddy magnetization''
terms.
The thermal
emission from the grain will be polarized.
Interstellar grains undergoing superthermal rotation will tend to
have their short axis $\hat{a}_1\parallel\bJ$, where $\bJ$ is the
angular momentum.
If $\bJ$ is in the plane of the sky, then
the degree of linear polarization will be
\beq
P= {I_e-I_h\over I_e+I_h} ~~~,
\label{eq:pol}
\eeq
\beq
I_e=
{1\over 2}\left[
{\epsilon_2  \over |1+L_2(\epsilon-1)|^2}
+
{\epsilon_2  \over |1+L_3(\epsilon-1)|^2}
\right]
+
\left[
{\mu_2       \over |1+L_1(\mu-1)|^2}
\right] ~~~,
\label{eq:I_e}
\eeq
\beq
I_h=
\left[
{\epsilon_2  \over |1+L_1(\epsilon-1)|^2}
\right]
+
{1\over 2}\left[
{\mu_2       \over |1+L_2(\mu-1)|^2}
+
{\mu_2       \over |1+L_3(\mu-1)|^2}
\right] ~~~.
\label{eq:I_h}
\eeq
From eq.\ (\ref{eq:pol}-\ref{eq:I_h}) it is apparent that (assuming
isotropic materials) polarization
arises only to the extent that {\it both}
(1) $|\epsilon-1|$ and/or $|\mu-1|$ differ from zero, {\it and}
(2) the geometrical factors $L_j$ differ from one another.

The contribution of magnetic dipole emission to the polarization from
Fe grains will be discussed in \S\ref{sec:polarization}

\section{Kramers-Kronig Relations\label{sec:KK}}

The Kramers-Kronig relations (see Landau \& Lifshitz 1960)
can be used to relate
total grain volume to the wavelength integral of extinction
(Purcell 1969), or to relate the wavelength-dependence of
linear and circular polarization produced by interstellar grains
(Shapiro 1975; Martin 1975a,b).
The original discussion
by Purcell (1969) assumed nonmagnetic grains; here we
generalize the discussion to include magnetic grains.
Following Purcell,
consider an electromagnetic plane wave propagating through the
dusty interstellar medium:
$\vec{E}={\rm Re}(\vec{E}_0 e^{ikx-i\omega t})$,
$\vec{H}={\rm Re}(\vec{H}_0 e^{ikx-i\omega t})$.
If $\tilde{\chi}_e(\omega)$ and
$\tilde{\chi}_m(\omega)$ are 
the electric and magnetic susceptibility of the
medium,
then the plane wave dispersion relation is
\beq
k^2c^2=\omega^2(1+4\pi\tilde{\chi}_e)(1+4\pi\tilde{\chi}_m) ~~~.
\eeq
For the interstellar medium we can assume $|\tilde{\chi}_e|\ll 1$,
$|\tilde{\chi}_m|\ll 1$.
Consider grains with number density $n_{gr}$, volume $V$,
and extinction cross section
$C_{ext}(\lambda)$, where $\lambda=2\pi c/\omega$.
Then
\beq
n_{gr} C_{ext}(\lambda) \approx
{8\pi^2\over \lambda}
{\rm Im}\left[ 
\tilde{\chi}_e(\omega) + \tilde{\chi}_m(\omega) 
\right] ~~~.
\label{eq:chi_vs_sigmaext}
\eeq
Similarly, the birefringence of the interstellar medium is proportional
to ${\rm Re}[\tilde{\chi}_e(\omega) + \tilde{\chi}_m(\omega)]$.
The linear response functions $\tilde{\chi}_e$ and $\tilde{\chi}_m$ must
separately satisfy the Kramers-Kronig relation,
\beq
{\rm Re}[\tilde{\chi_i}(\omega_0)] =
{2\over \pi}{\cal P}\!\int_0^\infty {\omega d\omega \over \omega^2-\omega_0^2}
{\rm Im}[\tilde{\chi_i}(\omega)] ~~~({\rm for}~ i=e,m) ~~~.
\eeq
where ${\cal P}$ indicates that the principal value is to be taken.
Taking $\omega_0=0$ one obtains
the generalization of Purcell's (1969) result:
\beq
{\rm Re}\left[\chi_e(0) + \chi_m(0)\right] = 
{c \over 4\pi^3} 
\int_0^\infty d\lambda ~n_{gr}{C_{ext}(\lambda)} ~~~.
\eeq
For randomly-oriented ellipsoids with semiaxes $a_1$, $a_2$, $a_3$,
we have
\beq
\tilde{\chi}_e (0) = {3V\over 4\pi}n_{gr} F(\epsilon_0; a_1/a_3, a_2/a_3)
\eeq
\beq
\tilde{\chi}_m (0) = {3V\over 4\pi}n_{gr} F(\mu_0; a_1/a_3, a_2/a_3)
\eeq
where
\beq
F(x) \equiv {1\over 9}\sum_{j=1}^3 {x-1 \over 1+(x-1)L_j}
\eeq
and the $L_j$ are the
geometric factors discussed above (see Table \ref{tab:geom_fact}).
For spheres, $L_j=1/3$ and $F(x)\approx1$ for $x\gg 1$
[Fig. 1 of Purcell (1969) shows $F(x)$ for
spheroidal grains ($a_1=a_2$ or $a_2=a_3$)].

Thus
\beq
n_{gr}V \left[ F(\epsilon_0) + F(\mu_0)\right] = {1\over 3\pi^2}
\int_0^\infty d\lambda ~ n_{gr}~C_{ext}(\lambda) ~~~,
\label{eq:KKresult}
\eeq
where $\epsilon_0$ and $\mu_0$ are the zero-frequency dielectric function
and magnetic permeability of the grain material.
For nonmagnetic materials, $F(\mu_0=1)=0$ and we recover Purcell's (1969)
result.  

Grain materials of interest have $\epsilon_0\gtsim 3$ (conducting grains
have $\epsilon_0\rightarrow\infty$).
For magnetic grains with $\mu_0\gtsim 3$,
$F(\mu_0)\approx F(\epsilon_0)$, and
we see from eq.\ (\ref{eq:KKresult}) that $\int C_{ext}d\lambda$
must be larger by a factor $[1 + F(\mu_0)/F(\epsilon_0)]$ than it
would have been had the grain been nonmagnetic.

Purcell (1979) used $\int n_{gr}C_{ext}d\lambda$ to estimate
$n_{gr}V$.
At first sight eq.\ (\ref{eq:KKresult}) 
appears to reduce the required
volume of interstellar dust by about a factor of two, but
it must be recognized that at the wavelengths $\lambda < 1 \mm$
which contributed to Purcell's estimate for $\int C_{ext}d\lambda$,
magnetic effects are indeed unimportant, so that

A second consequence, however, is that we see that 
if $F(\mu_0)\approx F(\epsilon_0)$, then magnetic dipole
absorption must contribute a cross section $C_{ext}^{(md)}$ such
that
\beq
\int C_{ext}^{(md)}d\lambda \approx 
\int C_{ext}^{(ed)}d\lambda ~~~;
\eeq
magnetic dipole absorption and emission {\it must} dominate the 
emission from (stationary) interstellar dust grains at the frequencies
$\nu\ltsim 30 \GHz$ where the grain material has a strong magnetic
response.

\section{Candidate Materials\label{sec:materials}}

In addition to ``ordinary'' paramagnetism, 
grains may exhibit strong magnetic ordering: either 
ferromagnetic,
ferrimagnetic, or antiferromagnetic
(see Morrish 1980, and \S\ref{sec:ferro} below).

Ordinary paramagnetism (\S\ref{sec:ordinary_para}) 
arises when the interaction producing
alignment of electron spins is just the interaction of the 
electron's magnetic moment with
the local magnetic field.
Ferromagnetism, ferrimagnetism, and antiferromagnetism 
(\S\ref{sec:ferro}) occur when the
electron spins are ordered due to 
the ``exchange interaction''.
By mass, Fe is the fifth most abundant element, following H, He, O, and C,
and nearly all of the interstellar Fe is in grains (Savage \& Sembach 1996).
Any strongly magnetic grain materials will
almost certainly
contain Fe.
Table \ref{tab:Feminerals} lists a number of possible Fe-containing
materials which are magnetically ordered at the $T\ltsim20\K$ temperatures
of interstellar dust.

Equations 
(\ref{eq:cabs_ed_ell},\ref{eq:cabs_md_ell}) show that the
absorption cross sections are proportional to the grain volume.
If 100\% of solar abundances (Anders \& Grevesse 1989) 
of Mg, Si, and Fe are incorporated into
silicates with the approximate composition 
Mg$_{1.1}$Fe$_{0.9}$SiO$_4$ and a density
$\rho\approx4\g\cm^{-3}$, the resulting volume of silicate per H atom is
$V_{0}=2.5\times10^{-27}\cm^3/\H$, which we will use as a fiducial
grain volume per H atom.

Suppose a fraction $f_{Y}$ of the Fe is incorporated into
a material $Y$ with density $\rho_Y$, 
with Fe contributing a fraction $z_Y$
of the mass.
Then the volume per H of material $Y$ is
\beq
{V_Y \over V_{0}} = 
0.30 f_{Y} \left( {4\g\cm^{-3}\over \rho_Y z_Y}\right) ~~~.
\label{eq:V_Y}
\eeq
Table \ref{tab:Feminerals} shows $V_Y/V_{0}$ for various candidate
minerals.
We see that Fe-rich materials are likely to make up a significant fraction
of the grain volume, and these materials tend to be magnetically-ordered.

%\placetable{tab:Feminerals}
%{\begin{center}[Editor: Place Table \ref{tab:Feminerals} here.]\end{center}}
\begin{table}
\begin{center}

\begin{tabular}{|c|c|r|c|c|}
\hline
	Mineral $Y$
	&Form\tablenotemark{b}
	&$4\pi M_s$(G)\tablenotemark{c}
	&$\rho_Y (\g\cm^{-3})$
	&$V_Y/V_{0}$\tablenotemark{d}	\\
\hline
Fe$_2$SiO$_4$ (fayalite)	&antiferromag.\tablenotemark{e}	&0	&4.39	&0.51$f_{Y}$ \\
FeSiO$_3$ (pyroxene)		&antiferromag.\tablenotemark{e}	&0	&	& \\
FeS (troilite)			&antiferromag.\tablenotemark{e}	&0	&4.83	&0.39$f_{Y}$ \\
FeCO$_3$ (siderite)		&antiferromag.\tablenotemark{e}	&0	&3.96	&0.63$f_{Y}$ \\
FeMgSiO$_4$ (olivine)		&antiferromag.\tablenotemark{e}	&0	&4.	&0.92$f_{Y}$ \\
$\alpha$Fe$_2$O$_3$ (hematite)	&antiferromag.\tablenotemark{e}	&0	&5.26	&0.33$f_{Y}$ \\
FeO (wustite)			&antiferromag\tablenotemark{e}.	&0	&5.75	&0.27$f_{Y}$ \\
Fe$_{1-x}$S (pyrrhotite)	&ferrimagnetic	&1130\tablenotemark{f}
							&4.6	&0.41$f_{Y}$ \\
MgFe$_2$O$_4$ (magnesioferrite)	&ferrimagnetic	&1760	&4.18	&0.51$f_{Y}$ \\
NiFe$_2$O$_4$ (trevorite)	&ferrimagnetic	&3770	&5.35	
	&0.47$f_{Y}$\tablenotemark{g} \\
$\gamma$Fe$_2$O$_3$ (maghemite)	&ferrimagnetic	&4780\tablenotemark{h}	&4.88	&0.36$f_{Y}$ \\
Fe$_3$O$_4$ (magnetite)		&ferrimagnetic	&6400	&5.2	&0.32$f_{Y}$ \\
Fe (iron)			&ferromagnetic	&22000	&7.86	&0.15$f_{Y}$ \\
\hline
\end{tabular}
\end{center}
\caption{\label{tab:Feminerals}
	Magnetic compounds containing Fe.\tablenotemark{a}
	}
\tablenotetext{a}{Data selected from review by Carmichael (1989).}
\tablenotetext{b}{Magnetic ordering at $T<20\K$ for ideal mineral.}
\tablenotetext{c}{$M_s$ is the spontaneous magnetization at $T<20\K$.}
\tablenotetext{d}{Volume 
	$V_Y$ per H atom of grain material $Y$ containing a
	fraction $f_Y$ of the solar abundance of Fe, relative to
	$V_{0}=2.5\times10^{-27}\cm^3/\H$,
	the volume of silicate which can be made using all the Si, Mg,
	and Fe from solar abundances.}
\tablenotetext{e}{Small 
	particles may be ferrimagnetic (see \S\ref{sec:antiferro}).}
\tablenotetext{f}{For $x=1/8$, where $M_s$ is maximized.}
\tablenotetext{g}{$f_Y<0.11$ required by cosmic abundances.}
\tablenotetext{h}{Dunlog \& \"Ozdemir 1997).}
\end{table}

\section{The Magnetic Susceptibility\label{sec:magnetic}}

Solids exhibit a variety of different responses to weak applied
oscillating magnetic fields.
When a static magnetic field is present, or the substance is
spontaneously magnetized, the susceptibility $\chi(\omega)$ has
a tensor nature
(see, e.g., Jones \& Spitzer 1967, \S IIIa).
In the present discussion we will treat $\chi$ as a scalar, but
will extend the discussion to spontaneously-magnetized material
in \S\ref{sec:Fe_single} and Appendix \ref{app:single_domain}.

Aside from the diamagnetism of superconductors and materials with
no unpaired electron spins, at low frequencies most materials are
characterized by $\chi(0)>0$, resulting
from either ``normal paramagnetism'' or magnetic ordering in the
form of superparamagnetism, ferrimagnetism, or ferromagnetism.
We consider these cases below, but first discuss the likely behavior
of the frequency-dependent susceptibility 
$\chi(\omega)\equiv\chi_1+i\chi_2$.

\subsection{Drude Susceptibility}

Suppose that the magnetization $M(t)$
obeys the equation of motion
\beq
\ddot{M}=\omega_0^2\left[\chi(0)H-M\right] -\dot{M}/\tau_0 ~~~,
\label{eq:eof_D}
\eeq
with three parameters: the static response $\chi(0)$,
a resonance frequency $\omega_0$,
and a characteristic damping time $\tau_0$.
Then we obtain
the Drude form for the susceptibility,\footnote{Morrish (1980)
	suggests a Lorentzian form for the susceptibility,
	corresponding to an equation of motion
	$\dot{M}=[\chi(0)H-M]/\tau$.
	Because this lacks an ``inertial'' term, we believe that it
	overestimates the magnetic response at very high frequencies.
	}
\beq
\chi(\omega) = 
{\chi(0) \over 1-(\omega/\omega_0)^2-i\omega\tau}
\label{eq:chi_res} ~~~,
\eeq
where we have defined
\beq
\tau \equiv (\omega_0^2\tau_0)^{-1} ~~~.
\label{eq:tau_def}
\eeq
The real and imaginary parts of $\chi(\omega)$ are
\beq
\chi_1 = \chi(0){1-(\omega/\omega_0)^2
\over
[1-(\omega/\omega_0)^2]^2 + (\omega\tau)^2}
\label{eq:chi1_res}
\eeq
\beq
\chi_2 = \chi(0){\omega\tau \over
[1-(\omega/\omega_0)^2]^2 + (\omega\tau)^2}~~~.
\label{eq:chi2_res}
\eeq
A susceptibility of the form (\ref{eq:chi_res}) arises, for example, 
as a solution to the
Bloch equations (c.f. Pake 1962, eq. 6-24) for a magnetized material.

The product $\omega_0\tau\equiv 1/\omega_0\tau_0$ 
determines the shape of $\chi(\omega)$.
For $\omega_0\tau<2$, $\chi(\omega)$ has a resonance near $\omega_0$.
For $\omega_0\tau>2$, the system is ``overdamped'', and
responds to a step function change in the applied $H$ with two distinct 
relaxation times.

%\placefigure{fig:chiforms}
%{\begin{center}[Editor: Place Fig.\ \ref{fig:chiforms} here.]\end{center}}
\begin{figure}
\epsscale{0.90}
\plotone{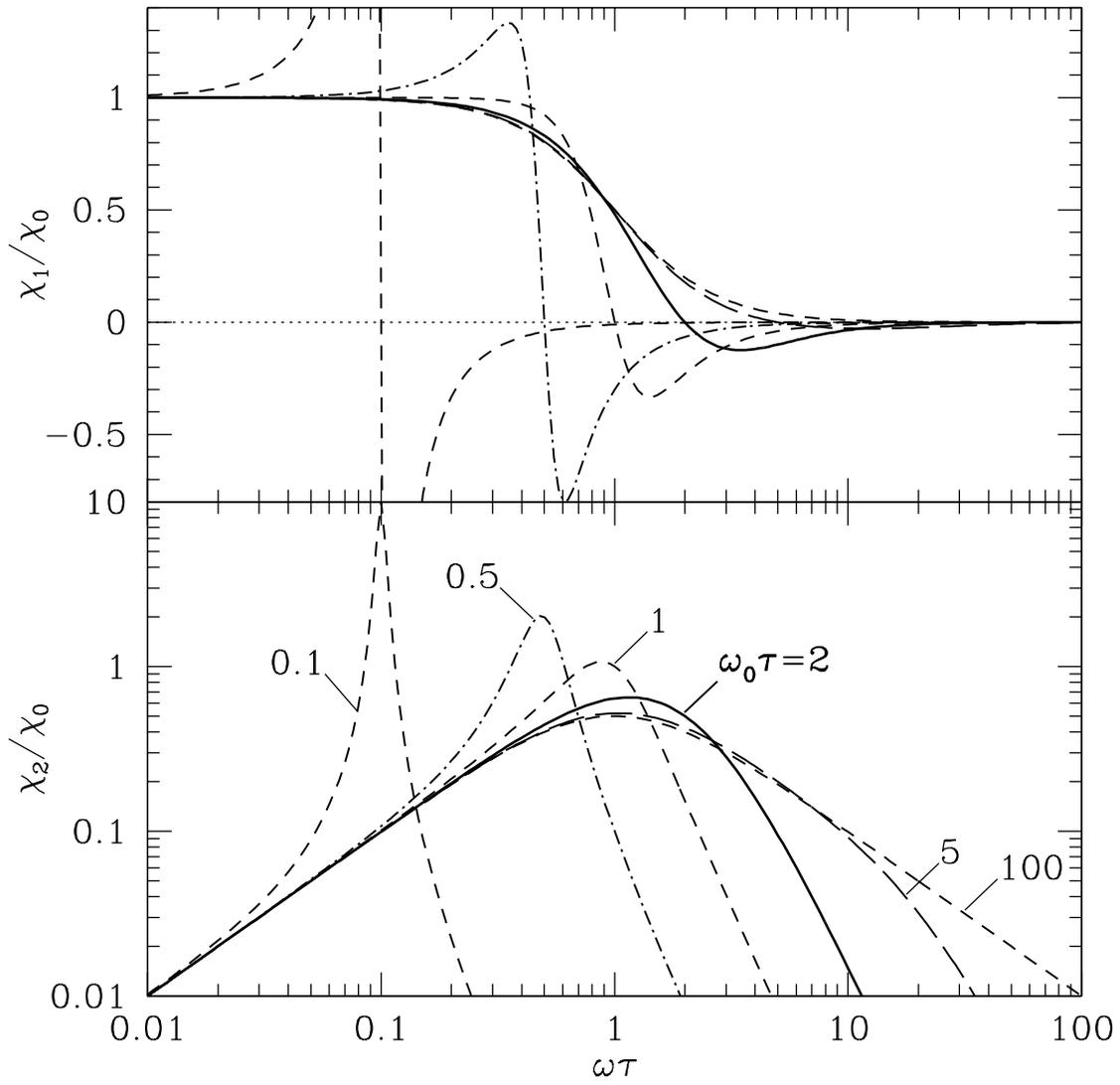}
\figcaption[f1.eps]{
	\label{fig:chiforms}
	Functional form considered for the magnetic susceptibility 
	$\chi=\chi_1+i\chi_2$, for different values of the
	parameter $\omega_0\tau$
	[see eq.\ (\protect{\ref{eq:chi_res}} -
	\protect{\ref{eq:chi2_res}})].
	}
\end{figure}

\subsection{``Critically-Damped'' Susceptibility \label{sec:critdamp}}

For $\omega_0\tau=2$, $\chi(\omega)$ is ``critically-damped'', with a single
relaxation time $\omega_0^{-1}=2\tau_0$.
We will consider this as a plausible form for the response
function $\chi(\omega)$ for a system which is essentially nonresonant.
For this case $\chi$ takes the simple form, shown as the heavy
curves in Fig.\ \ref{fig:chiforms}:
\beq
\chi^{(cd)}={\chi(0)\over (1-i\omega\tau/2)^2} ~~~,
\label{eq:chi_cd}
\eeq
\beq
\chi_1^{(cd)} = 
\chi(0){1-(\omega\tau/2)^2\over \left[1+(\omega\tau/2)^2\right]^2}
\label{eq:chi_cd1}~~~,
\eeq
\beq
\chi_2^{(cd)} = 
\chi(0){\omega\tau\over \left[ 1+(\omega\tau/2)^2\right]^2}
\label{eq:chi_cd2} ~~~.
\eeq
The superscript ``(cd)'' stands for ``critically-damped''.
We will use $\chi^{(cd)}$ to estimate the response associated
with the magnetization of a single-domain particle, or of 
paramagnetic materials.
We will identify the characteristic frequency $\omega_0=1/2\tau_0$ with
the gyrofrequency of an electron in the internal field.
From eq. (\ref{eq:chi_cd}-\ref{eq:chi_cd2}) we see that the response
falls off rapidly when the frequency $\omega > \omega_0=2/\tau$.

\section{Paramagnetism\label{sec:paramag}}

\subsection{Ordinary Paramagnetism\label{sec:ordinary_para}}

For ordinary paramagnetism,
the zero-frequency susceptibility is (Draine 1996)
\beq
\chi(0) \approx 4\times10^{-2}f_p\left({p\over 5.5}\right)^2 
\left({15\K\over T}\right) ~~~,
\eeq
where $f_p$ is the fraction of the atoms which are paramagnetic, with
magnetic moments $p\mu_B$, where the Bohr magneton $\mu_B\equiv e\hbar/2m_ec$.
If essentially all of the interstellar 
Mg, Fe, and Si are incorporated into material
with approximately the composition of MgFeSiO$_4$, then a fraction
$f_p\approx 1/7$ of the atoms would be Fe, presumably in the
form of Fe$^{2+}(^5D_4)$ or Fe$^{3+}(^6S_{5/2}$) ions;
these have
$p\approx 5.4$ and 5.9, respectively (Morrish 1980).
Fayalite (Fe$_2$SiO$_4$) is antiferromagnetic (Carmichael 1989) and
it appears that intermediate olivines (Mg$_x$Fe$_{2-x}$SiO$_4$) will
also be antiferromagnetic (Duff 1968).
The magnetic character of amorphous olivine
is uncertain.  While weak ferrimagnetism seems likely
(see below), here we consider paramagnetic behavior.

\begin{figure}
\epsscale{0.90}
\plotone{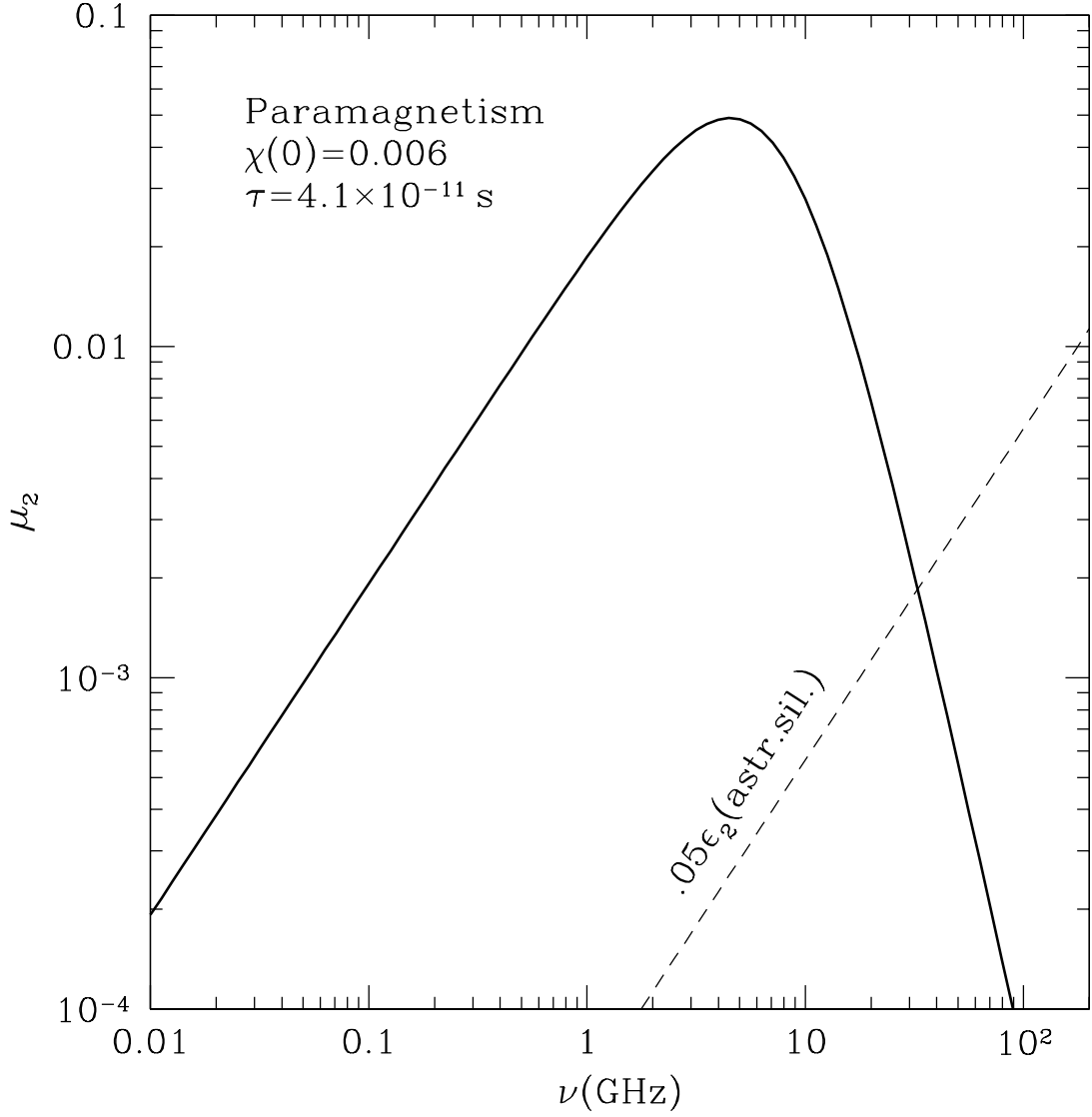}
\figcaption[f2.eps]{
	\label{fig:PM}
	Estimated imaginary component of magnetic permeability for
	``normal'' paramagnetism in silicate grains (curve labelled
	PM).
	The dot-dash line labelled ``$.05\epsilon_2$'' shows the level
	above which magnetic dipole absorption due to paramagnetism 
	dominates electric
	dipole absorption (see eq.\ \protect{\ref{eq:cabsmd_over_cabsed}}).
	}
\end{figure}

The damping time $\tau$ is the spin-spin relaxation time, which is essentially
the time for electron precession in the random magnetic fields within
the solid (see Caspers 1967).
For amorphous olivine 
we estimate the r.m.s. gyrofrequency to be (Draine 1996)
\beq
\omega_0/2\pi\approx 8 \GHz ~~~.
\eeq
Draine (1996) used a different functional form\footnote{
	The form used by Draine (1996) had $\chi_2 \propto \omega^{-1}$
	for $\omega\rightarrow\infty$.
	}
to estimate
the paramagnetic susceptibiity, but we now consider
the ``critically-damped'' form $\chi^{(cd)}$ 
[eq.(\ref{eq:chi_cd})]
to provide a better estimate at frequencies $\omega \gg \omega_0$.
In Fig.\ \ref{fig:PM}
we show 
$\mu_2=4\pi\chi_2^{(cd)}$,
for 
$\tau = 2/\omega_0 = 4\times10^{-11}\s$.

%\placefigure{fig:PM}
%{\begin{center}[Editor: Place Fig.\ \ref{fig:PM} here.]\end{center}}

At frequencies $\nu\ltsim200$GHz ``astronomical silicate'' 
(Draine \& Lee 1984)
has $\epsilon_1\approx 11.6$, 
$\epsilon_2\approx1.13\times10^{-3}(\nu/{\rm GHz})$.
From eq.\ (\ref{eq:cabsmd_over_cabsed}), it can then be seen 
that $C_{abs}^{(md)}>C_{abs}^{(ed)}$
when $\mu_2 > .05\epsilon_2$
(assuming $\mu_1\approx1$, as is the case for ordinary paramagnetism), 
so for reference we also show $.05\epsilon_2$ in
Fig.\ \ref{fig:PM}.
From Fig.\ \ref{fig:PM} it can then be seen
that if FeMgSiO$_4$ grains are merely paramagnetic,
the thermal emission from these grains
at
$\nu\ltsim 30$ GHz
will evidently be dominated by 
magnetic dipole emission arising from 
thermal fluctuations in the magnetization.

\subsection{Stark Splitting in Paramagnetic Materials}

Additional magnetic absorption can arise in paramagnetic
material 
when crystalline electric fields produce Stark-effect splitting of
the magnetic sublevels, and this splitting corresponds to
microwave frequencies (Kittel \& Luttinger 1948).
Electric dipole transitions are forbidden, since the sublevels have the
same parity, but magnetic dipole transitions are allowed.

This type of absorption has been studied in the
so-called
Zero Field Resonance experiments (see Bramley \& Strach 1983 and references
therein).
For Fe$^{3+}$ resonances were observed at $24.2$ and
$35.4$~GHz in natural amethyst and at $7.1$ and $8.8$ GHz
for synthetic brown quartz. Similar
measurements in sapphire revealed
resonances at $11.8$ and $18.9$ GHz. 
%A series of resonances
%in the range from $1$ to $4$~GHz was reported for Mn$^{2+}$
%in Mg(NH$_4$)$_2$(SO$_4$)$_2$6H$_2$O  and in the range
%from 6.5 to 11.5~GHz for the same ion in NH$_4$Cl. 

For pure single crystals the absorption is
localized in lines, while a continuum
of absorption features is expected for 
impure and noncrystalline interstellar grains.
The strength of this absorption can be estimated
(see Appendix \ref{app:stark_effect}).
At the temperatures of interstellar grains, it appears that
a permeability $\mu_2\ltsim10^{-3}$ would result.
We see from Fig. \ref{fig:PM} that this would be
a minor contribution to grain absorption and emissions 
compared even with what we estimate for
``ordinary'' paramagnetism at $1 - 30\GHz$,
and far below what we estimate in \S\ref{sec:ferro} 
for ferromagnetic or ferrimagnetic
materials at microwave frequencies.
Accordingly, as a continuum emission process this mechanism appears
to be unimportant.

The absorption due to Stark splitting could be concentrated
in a few absorption features, rather than being spread into a continuum.
Even so, the effect will be weak.
If, say, 5\% of the atoms had Stark splitting features within a
frequency range $\Delta\ln\omega=0.01$, then 
we see from eq. (\ref{eq:mu_stark}) that
$\mu_2\approx 0.14 (|\mu_{ul}|^2/\mu_B^2)$ within the narrow feature.
At, say, 5 GHz this is only a few times larger than the value of
$\mu_2$ which we estimate for ordinary paramagnetism 
(see Fig.\ \ref{fig:PM}) and hence would result in a narrow emission feature
with central intensity only a few times what we estimate for 
the continuum due to ordinary
paramagnetism, which we shall see below is itself 
two orders of magnitude weaker
than the observed continuum emission.
Nevertheless we cannot exclude the possibility that future sensitive
measurements of the microwave background might disclose weak spectral
features arising from the Stark effect.

\subsection{Magnetic Materials in Interstellar Grains\label{sec:mag_grains}}

As noted in \S\ref{sec:materials},
the substantial fraction of Fe in interstellar grains creates the
possibility that some fraction of the grains could be 
magnetic, with
the ``exchange interaction'' resulting in ordering
of the atomic magnetic moments within a single magnetic domain
(see Morrish (1980) for a review of the physics of magnetism, and
Dunlop and \"Ozdemir (1997) for an excellent review of magnetic minerals).

Ferromagnetic materials as components of interstellar grains were
apparently first considered by
Spitzer \& Tukey (1951), who discussed the possible formation of ferromagnetic
materials in interstellar grains as the result of grain-grain collisions: 
metallic iron,
Fe$_3$O$_4$,
$\gamma$Fe$_2$O$_3$, and MgFe$_2$O$_4$ were considered likely products.
Jones \& Spitzer (1967) suggested that interstellar grains might contain
very small clusters of magnetic materials such as Fe$_3$O$_4$ or
$\gamma$Fe$_2$O$_3$.
Shapiro (1975) proposed that the observed polarization of starlight
could be produced by platelets of Fe$_3$O$_4$.
Sorrell (1994) presented a model for the origin of small Fe$_3$O$_4$ clusters
in H$_2$O ice mantles irradiated by cosmic rays.
Ferromagnetic properties of mixed MgO-FeO-SiO grains were discussed by
Duley (1978).

Bradley (1994)
argued that certain interplanetary dust particles
(``Glass with Embedded Metals and Sulfides'', or ``GEMS'')
consisted of aggregates of $\sim0.1\micron$ interstellar grains, with
nanometer-sized Fe-Ni metal inclusions.
Martin (1995) argued that the properties of these grains were consistent
with being interstellar dust, and that the
Fe-Ni inclusions 
could be superparamagnetic and able to bring about
alignment with the galactic magnetic field.
Goodman \& Whittet (1995) noted that the numbers of such
inclusions were consistent with the requirements of Mathis' (1986)
hypothesis to account for the dependence of degree of alignment on
grain size.

The S in GEMS appears to be in the form of FeS; if the excess of Fe over
S is metallic, then the
atomic abundances given by Bradley (1994) for 3 bulk GEMS
indicates a volume filling factor $\phi\approx0.03$
for metallic Fe-Ni;
%[Martin (1995) estimated $\phi\approx0.02$]; 
we will use this value when
discussing the properties of silicate grains with metallic inclusions
(\S\ref{sec:Fe_incl}).

\section{Ferromagnetic and Ferrimagnetic Materials\label{sec:ferro}}

There are three classes of magnetically-ordered materials.
In {\it ferromagnetic} materials the atomic spins within a domain are parallel;
in {\it ferrimagnetic} and {\it antiferromagnetic} materials the magnetic 
ions are located on two magnetic sublattices of oppositely 
directed spins and magnetic moments.
In {\it antiferromagnetic} materials the magnetic moments of the
two sublattices are exactly opposite in direction and magnitude, so the
net magnetization is zero (for a perfect crystal at zero temperature).
In {\it ferrimagnetic} materials the magnetic sublattices do not exactly
compensate, either because the magnetic moments of the two sublattices
differ in magnitude
(normal ferrimagnetism) or are not precisely opposite in direction
(spin-canted ferrimagnetism).

\subsection{Single-Domain Iron \label{sec:Fe_single}}

Applied magnetic fields change the magnetization of a bulk 
sample by two processes:
rotation of the magnetization within a domain, and motion of domain walls.
Grains smaller than a critical radius $a_c$ will contain only a single domain; 
for Fe
$a_c\approx 3\times 10^{-6}\cm$ (Morrish 1980).
Such single-domain behavior can take place for clusters of materials
which are ferromagnetic (e.g., Fe) or ferrimagnetic (e.g., Fe$_3$O$_4$,
or $\gamma$Fe$_2$O$_3$).
Spontaneous magnetization has been observed for Fe clusters as small as
$N_{cl}=20$ atoms (Billas, Ch\^atelain \& de Heer 1994) 

If the magnetic material consists of a single domain, then 
there is clearly no motion of domain walls involved in the magnetization
process, unlike the case for bulk samples, so susceptibilities measured
for bulk samples are inapplicable.

When a ferromagnetic or ferrimagnetic 
material consists of a single domain, it will be spontaneously
magnetized with magnetization ${\bf M}_s$ along one of the ``easy'' directions.
If a weak oscillating field ${\bf H}$ 
is now applied, the single-domain sample will
exhibit a susceptibility which depends on the direction of ${\bf H}$.
In Appendix \ref{app:single_domain}
we show that at low temperatures only the component 
of ${\bf H}$
perpendicular to the spontaneous 
magnetization induces a change in magnetization of
the sample: $\Delta {\bf M} = \chi_\perp {\bf H}_\perp$.
For Fe, we estimate $\chi_\perp(0)\approx0.3\chi_{bulk}(0)$ 
(see Appendix \ref{app:single_domain}).

\begin{figure}
\epsscale{0.90}
\plotone{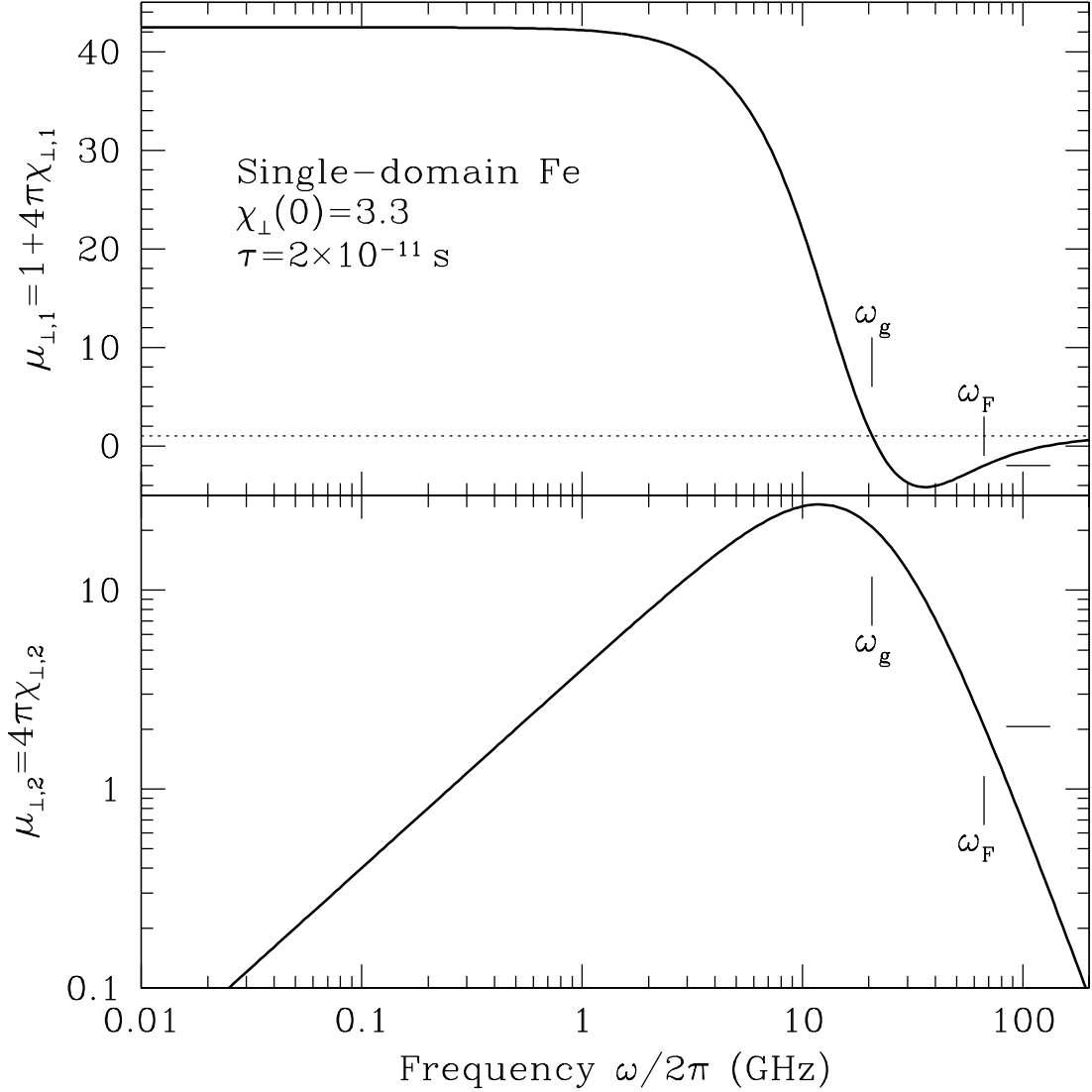}
\figcaption[f3.eps]{
	\label{fig:Fe_single}
	Real and Imaginary parts of the permeability $\mu=1+4\pi\chi_\perp$
	for single-domain Fe with ${\bf H}$ perpendicular to the direction
	of spontaneous magnetization
	(see text).
	The estimated gyrofrequency $\omega_g$ (where $\mu_{\perp,1}=1$) 
	and Fr\"ohlich frequency 
	$\omega_F$ ($\mu_{\perp,1}=-2$) are indicated.
	}
\end{figure}

Since changing the magnetization of the sample involves only small
reorientations of the magnetic moments, and no motions of domain walls,
we assume that the only frequency characterizing the 
response is the gyrofrequency
$\omega_g=(e/m_ec)(4\pi M_s/3)$ of a magnetic moment with $g\approx2$ in the
internal field $4\pi M_s/3$ to which each of the dipoles is subject.
For Fe, with $4\pi M_s=22$kG, we estimate
$\omega_0/2\pi\approx 20\GHz$.
We use $\chi^{(cd)}$ [eq.(\ref{eq:chi_cd})] 
with
$\tau\approx2/\omega_0\approx 1.6\times10^{-11}\s$
to estimate $\chi_\perp(\omega)$
(see Fig.\ \ref{fig:Fe_single}).

%\placefigure{fig:Fe_single}
%{\begin{center}[Editor: Place Fig.\ \ref{fig:Fe_single} here.]\end{center}}

The
energy dissipation rate is proportional to ${\rm Im}(\mu_\perp) H_\perp^2$.
For isotropic radiation incident on the grain,
$\langle H_\perp^2\rangle = (2/3)\langle H^2\rangle$.
Assuming the higher order term in eq.\ (\ref{eq:cabs_md})
to be negligible,
we would therefore take the angle-averaged magnetic dipole absorption
cross section for randomly-oriented spheres to be
\beq
\langle C_{abs}^{(md)}\rangle \approx 
V{\omega\over c}
{6\mu_{\perp,2}\over (\mu_{\perp,1}+2)^2+\mu_{\perp,2}^2}
\label{eq:cabsperp}
\eeq
whereas for randomly-oriented ellipsoids with semiaxes
$a_1\leq a_2\leq a_3$, magnetized along the long axis,
\beq
\langle C_{abs}^{(md)}\rangle \approx
{V\omega\over 3c}
\sum_{j=1}^2
{\mu_{\perp,2}\over L_j^2\left[(\mu_{\perp,1}-1+L_j^{-1})^2+\mu_{\perp,2}^2
\right]}
\label{eq:cabsperp_ell}
\eeq
with $\mu_\perp=1+4\pi\chi_\perp$ obtained using eq. (\ref{eq:chi_cd})
with either $\chi_\perp(0)$ obtained from eq. (\ref{eq:chi_perp_100}).

\subsubsection{Fr\"ohlich Resonance Condition}

We now observe that if our estimate for $\chi_\perp(\omega)$ in
Fig.\ \ref{fig:Fe_single} is correct,
then Fe particles will have the magnetic analogue of
a Fr\"ohlich resonance (Fr\"ohlich 1949; Bohren \& Huffman 1983) -- a
peak in $C_{abs}$ where $\mu_{\perp,1}=1-L_j^{-1}$, thus minimizing the
denominator in eq.\ (\ref{eq:cabs_md_ell}) 
or in eq.\ (\ref{eq:cabsperp_ell}).

It is apparent that for the functional forms in Fig.\ \ref{fig:chiforms}
there are either zero or two frequencies where 
this condition is satisfied; if there are two,
then
$C_{abs}$ peaks at the higher of the two frequencies, which we denote
$\omega_F$.

The Fr\"ohlich resonance condition for spheres 
requires that $\mu_1 < -2$ for some
range of frequencies, or $\chi_1 < -3/4\pi$.\footnote{
	\label{fn:Frohlich}
	For $\chi^{(cd)}(\omega)$ [eq.\ (\ref{eq:chi_cd})]
	it is easy to show that the
	Fr\"ohlich resonance condition for spheres, $\mu_1=-2$, 
	can be satisfied provided $\chi(0) \geq 6/\pi = 1.91$,
	with $\omega_F\geq \surd 3 \omega_0$.
	However, $\omega_F$
	is associated with a distinct peak in $C_{abs}$ only for
	$\chi(0) \gtsim 3$.
	}
Whether or 
not Fe particles will have a Fr\"ohlich resonance will depend on the
details of the frequency-dependence of $\chi$, which at this time
is not experimentally-determined.
We are therefore forced to rely on model-dependent estimates.
With our estimate of $\chi_\perp$ (see Fig.\ \ref{fig:Fe_single}), 
we estimate that, for spheres, $\omega_F/2\pi\approx70\GHz$.
For ellipsoids with moderate axial ratios, the resonance will be
shifted slightly.
For example, for a 1:1.5:2 ellipsoid, magnetized along the long axis,
the Fr\"ohlich resonances along axes 1 and 2 occur for
$\mu_1=-1.07$ and $-2.28$, or $\omega_{F}/2\pi\approx100\GHz$ and $60\GHz$.

Recalling the Kramers-Kronig relation (Landau \& Lifshitz 1960)
\beq
\chi_1(\omega) = {2\over\pi} {\cal P}\!\int_0^\infty {x\chi_2(x)\over x^2-\omega^2} dx
\label{eq:KK}
\eeq
(where ${\cal P}$ denotes principal value integral)
and the fact that $\chi_2 >0$, one sees that the integrand is 
negative for $x < \omega$.
Quite generally, $\chi_2(\omega)$ will have $\chi_2(0)=0$,
will increase with increasing $\omega$ (linearly at low frequencies), peak,
and then eventually decline to zero
at high frequencies.
If $\chi_2$ declines more rapidly than $\omega^{-2}$ at high frequencies,
then we expect to have $\chi_1<0$ above some frequency.
Because metallic Fe has 
$\chi_0 \gg 1$, it appears likely that there will be a frequency
range where $4\pi\chi_1<-3$, so that the Fr\"ohlich resonance condition 
can be satisfied.

The functional form $\chi^{(cd)}$ used here provides what we consider a
reasonable estimate for $\chi_\perp$, but the predicted Fr\"ohlich frequency
$\omega_F/2\pi\approx70\GHz$ is model-dependent and 
should be regarded as quite uncertain.

\subsection{Bulk Iron\label{sec:Fe_bulk}}

The static susceptibility of bulk, metallic Fe is $\chi(0)\approx12$.
The susceptibility of Fe as a function of frequency has been measured
using fine wires,
thin lamina, and powders (Allanson 1945; Epstein 1954) up to 
$\sim10\,$GHz.
The response of bulk material will result from motion of domain walls
as well as the deflections of magnetization within single domains
as discussed above.
We treat these two responses as additive:
$\chi({\rm bulk\,Fe})=(2/3)\chi_\perp+\chi_{dw}$,
where $\chi_{dw}$ is the contribution from motion of domain walls.

\begin{figure}
\epsscale{0.90}
\plotone{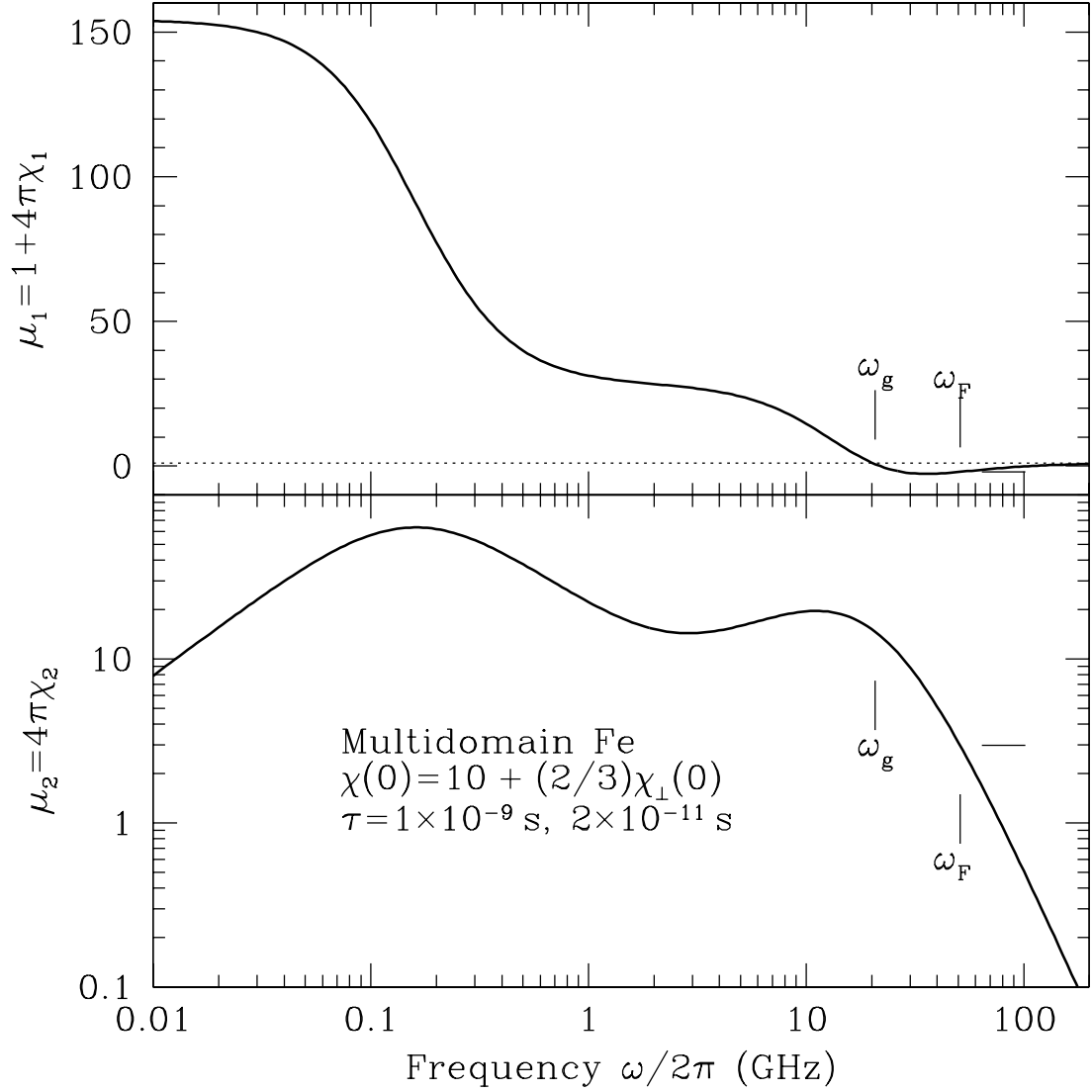}
\figcaption[f4.eps]{
	\label{fig:Fe_bulk}
	Real and imaginary permeability $\mu$
	for multidomain Fe (see text).
	The gyrofrequency $\omega_g$ and Fr\"ohlich frequency $\omega_F$
	are shown.
	}
\end{figure}

The experimental results show considerable scatter, 
but we can reproduce
the general trends if $\chi_{dw}$ is given by
eq.(\ref{eq:chi_res}) with $\chi(0)\approx 10$,
$\tau\approx1\times10^{-9}\s^{-1}$, and $\omega_0\approx 10^{10}\s^{-1}$
In Fig.\ \ref{fig:Fe_bulk} we plot $\mu_2$ estimated for multidomain 
metallic Fe using these parameters.

%\placefigure{fig:Fe_bulk}
%{\begin{center}[Editor: Place Fig.\ \ref{fig:Fe_bulk} here.]\end{center}}

\subsection{Fe Inclusions \label{sec:Fe_incl}}

Consider spherical single-domain ferromagnetic inclusions with 
volume filling factor
$\phi$ distributed in a nonmagnetic matrix.
Suppose each randomly-oriented 
inclusion to be spontaneously magnetized along one of its
``easy'' directions.
We may suppose that,
in effect, (2/3) of these inclusions have their magnetization ${\bf M}$ 
perpendicular
to the direction of the applied ${\bf H}$, with the remaining 1/3 either parallel
or antiparallel to ${\bf H}$ (and therefore not contributing to the
susceptibility).

\begin{figure}
\epsscale{0.90}
\plotone{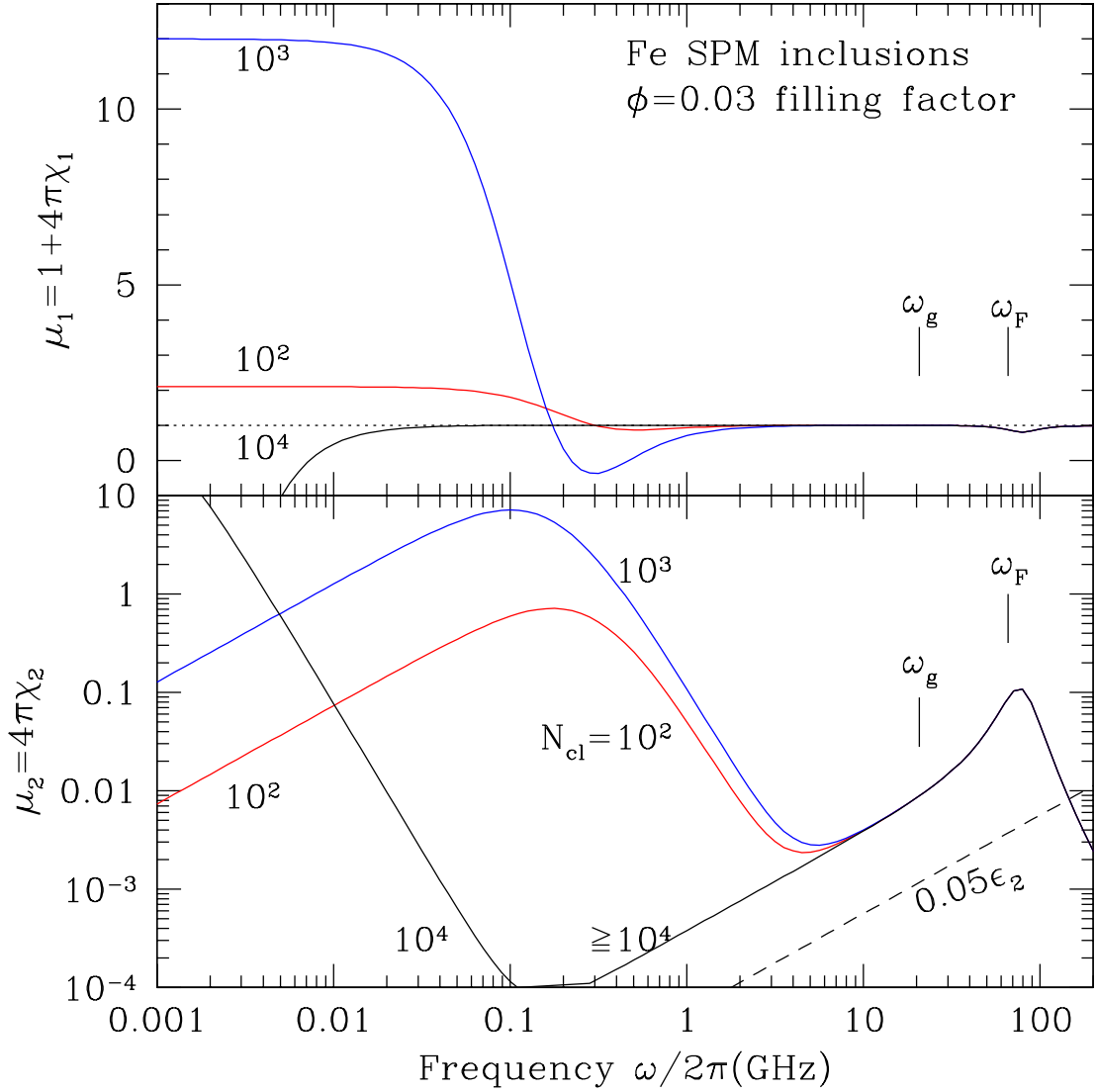}
\figcaption[f5.eps]{
	\label{fig:SPM}
	Same as Fig.\ \protect{\ref{fig:Fe_bulk}} but for
	material with 3\% of the volume due to 
	inclusions consisting of $N_{cl}$ Fe atoms.
	The Fr\"ohlich frequency $\omega_F$ is that estimated for
	single-domain Fe (see text).
	}
\end{figure}

The effective
susceptibility of the composite material
may be estimated using effective medium theory (Bohren \& Huffman 1984).
For the present case of spherical inclusions with small volume filling factor
$\phi \ll 1$,
Maxwell-Garnett effective
medium theory is appropriate, so we estimate
\beq
\chi_{\rm eff} = {(2/3)\phi\chi_\perp
\over
1+(4\pi/3)\chi_\perp(1-2\phi/3)}
\label{eq:largeinclusions}
\eeq
where $\chi_\perp$ is given by 
$\chi^{(cd)}$ [eq.(\ref{eq:chi_cd})].
In Fig.\ \ref{fig:SPM} we show $\mu$ estimated for Fe inclusions with a
volume filling fraction $\phi=0.03$ (the estimated filling fraction
of Fe-Ni inclusions in GEMS; see \S\ref{sec:mag_grains} 
in a nonmagnetic (e.g., silicate)
medium.
Note the peak in the effective $\mu_2$ for the composite medium at
$\omega/2\pi\approx 70\GHz$; this arises because of the Fr\"ohlich
resonance in the individual single-domain Fe inclusions.

%\placefigure{fig:SPM}
%{\begin{center}[Editor: Place Fig.\ \ref{fig:SPM} here.]\end{center}}

\subsection{Superparamagnetism \label{sec:superpara}}

If the Fe atoms are aggregated into single-domain clusters which are
sufficiently small, then thermal fluctuations will cause the
magnetization to fluctuate in direction, 
with the most probable magnetization directions
being the most energetically favorable ones.
If the inclusions are sufficiently small
($N_{cl}\ltsim 6\times10^5$ for Fe at room temperature),
the energy barrier associated with reorientation of 
the magnetization is small enough
that thermal fluctuations can reorient the magnetization of 
a single particle on time scales of seconds or less.
Such clusters are termed {\it superparamagnetic}.
If a field $H$ is applied, the change in magnetization will be
greater than would be the case for larger clusters, where only a small
deflection in the direction of magnetization takes place, as discussed
in \S\ref{sec:Fe_single}

Suppose that the grain consists of a nonmagnetic (or weakly magnetic)
material containing 
superparamagnetic clusters of $N_{cl}$ Fe atoms,
with volume filling factor $\phi_{sp}$.
A single grain then contains 
$N\approx 3.5\times10^8 \phi_{sp}N_{cl}^{-1}(a/10^{-5}\cm)^3$ clusters.

The susceptibility of the medium may be approximated as
\beq
\chi\approx 
{(2/3)\phi_{sp}\chi_\perp(0) \over
1 + (4\pi/3)\chi_\perp(0)\left[1-(2/3)\phi_{sp}\right]} 
+
\chi_{sp} ~~~.
\eeq
The first term is what would be expected if the individual clusters have
their magnetization locked into the ``easy'' direction, with a small
reorientation of the magnetization resulting from the transverse component
of the applied ${\bf H}$ [see eq.\ (\ref{eq:largeinclusions})].
The second term is what is expected from alignment of individual clusters
assuming them to be free to orient their magnetization in any direction.
We note that some previous discussions of the susceptibility of 
superparamagnetic materials (e.g., Jones \& Spitzer 1967;
Lazarian 1995; Draine 1996)
have omitted the first term when considering superparamagnetism.
In principle, these two estimates should not be treated as simply additive,
but in general one or the other will dominate, and 
factor-of-two accuracy is sufficient at this time.

The superparamagnetic contribution at zero frequency is estimated to be
\beq
\chi_{sp}(0)
\approx
.035 \phi_{sp} N_{cl}\left({15\K\over T}\right) ~~~.
\eeq
In general (even if spherical), a cluster must overcome
an energy barrier in order to substantially reorient its magnetization.
Laboratory experiments show the relaxation 
process to be be thermally activated, with a characteristic relaxation rate 
\beq
\tau^{-1}\approx A \exp(-N_{cl}\theta/T)
\label{eq:nu0}~~~,
\eeq
with $A\approx10^9\s^{-1}$ and
$\theta\approx0.011K$ for metallic Fe
spheres (Bean \& Livingston 1959; Jacobs \& Bean 1963).
We will employ (\ref{eq:chi_res}) with $\omega_0=10^{10}\s^{-1}$
to estimate $\chi_{sp}(\omega)$.

Fig.\ \ref{fig:SPM} shows $\mu_2$ for 
silicate grains containing superparamagnetic Fe clusters 
with volume filling factor $\phi_{sp}=0.03$.
Such inclusions, if present in all of the silicate grains, would
account for $f=\phi_{sp}/0.15=20\%$ 
of the total Fe abundance (see Table \ref{tab:Feminerals}).
Permeabilities are shown for $N_{cl}=10^2, 10^3$, and $10^4$ Fe atoms per
cluster.

The superparamagnetic response is large at low frequencies, but is
small compared to normal paramagnetism at frequencies $\nu\gtsim 1$ GHz,
because the estimated relaxation rate is too slow; at frequencies
$\nu\gtsim 1$ GHz, the magnetic response is essentially that of the
individual domains having their spontaneous magnetizations deflected slightly
by the transverse component of the magnetic field.

We conclude that while superparamagnetism may lead to 
enhancements of magnetic dissipation at $\nu\ltsim 1$ GHz, at higher
frequencies it appears unlikely to make an appreciable 
contribution to absorption
by interstellar grains.
Therefore the high-frequency response of grains with Fe inclusions 
depends only on the volume fraction $\phi_{sp}$ of Fe inclusions, but is
insensitive to the number $N_{cl}$ of Fe atoms per inclusion,
provided only that $N_{cl}\gtsim20$ so that the inclusions are ferromagnetic.

\subsection{Ferrimagnetic Minerals: Magnetite and Maghemite\label{sec:ferri}}

%Much of the available
%data on magnetic behavior at GHz frequencies 
%concerns various ferrites\footnote{
%	Ferrites are ferrimagnetic materials which are also
%	nonconducting.
%	The study of high frequency magnetic response in magnetic
%	materials has concentrated on
%	ferrites because of their low
%	eddy current losses.
%	}
%because of their microwave engineering applications
%(Lax \& Button 1962; Helszajn 1969).
%In some cases the material may
%exhibit some degree of resonant behavior.
%For example, the ``sintered Ni$_2$Y ferrite'' of Shin \& Oh (1993)
%is characterized by eq.\ (\ref{eq:chi_res}) with
%$\chi(0)=0.2$, $\omega_0\approx5\times10^{10}\s^{-1}$,
%%\tau\approx 3\times10^{-11}\s$,
%and $\omega_0\tau\approx 0.7$.
%Unfortunately, 
%it is unclear
%to what extent these data for ``exotic'' materials 
%can be used for estimation of the properties of
%interstellar dust.

Magnetite (Fe$_3$O$_4$) is a commonly-occuring terrestrial mineral, and
is a plausible interstellar grain material.
The low temperature spontaneous magnetization $M_s\approx 480\G$,
from which we estimate a gyrofrequency 
$\omega_g/2\pi\approx 6\GHz$.

At room temperature magnetite is cubic, and the ``easy'' direction is
$\langle 111\rangle$.
At $\sim120\K$ it undergoes a phase transition with a slight
distortion of the unit cell from cubic to monoclinic symmetry; near this
temperature the magnetic anisotropy parameter $K_1$ (see
Appendix \ref{app:single_domain}) changes sign and the ``easy''
direction changes to $\langle 100\rangle$.
The magnetic anisotropy parameters have been measured down to
$77\K$ (Syono \& Ishikawa 1963).
Extrapolation to $T\approx18\K$ is uncertain; we take
$K_1\approx 1.5\times10^5\erg\cm^{-3}$,
from which we estimate $\chi_\perp(0)\approx0.83$ using
eq.\ (\ref{eq:chi_perp_100}).

\begin{figure}
\epsscale{0.90}
\plotone{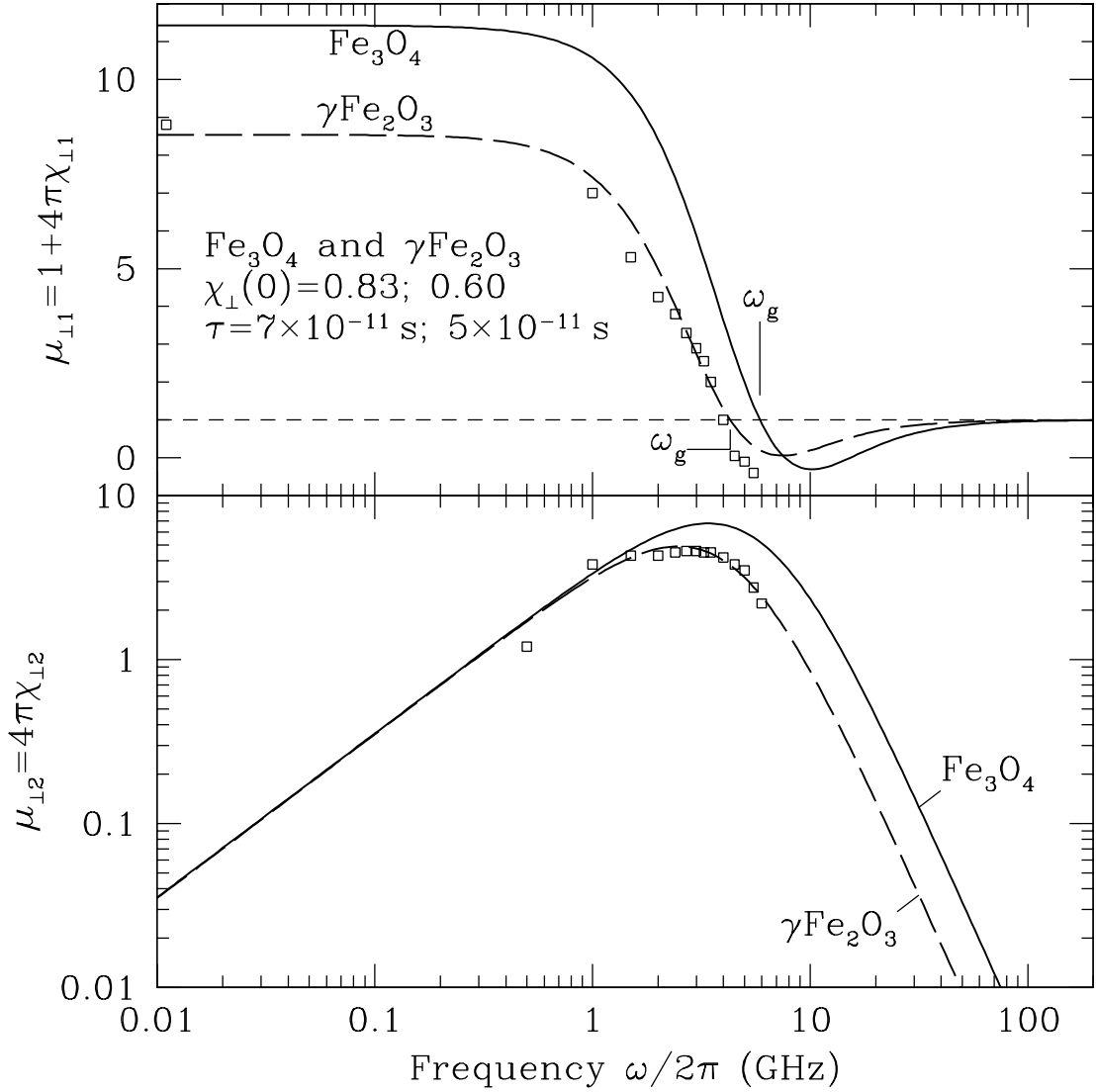}
\figcaption[f6.eps]{
	\label{fig:Fe_3O_4_single}
	Same as Fig.\ \protect{\ref{fig:Fe_single}} but for
	single-domain magnetite (Fe$_3$O$_4$), 
	an example of a ferrimagnetic material.
	Also shown are experimental results for $\gamma$Fe$_2$O$_3$
	from Valsytn et al.\ (1962).
	}
\end{figure}

We have not found data on the high frequency susceptibility of magnetite.
Taking $\tau=2/\omega_g=5\times10^{-11}\s$,
we estimate $\mu$ for single-domain Fe$_3$O$_4$ using
$\chi^{(cd)}$ [eq.\ (\ref{eq:chi_cd})]; the results are
shown in Fig.\ \ref{fig:Fe_3O_4_single}.

%\placefigure{fig:Fe_3O_4_single}
%{\begin{center}[Editor: Place Fig.\ \ref{fig:Fe_3O_4_single} here.]\end{center}}

The low temperature oxidation product of magnetite is maghemite,
$\gamma$Fe$_2$O$_3$,
for which we estimate $\omega_g/2\pi\approx 4\GHz$.
From $M_s$ and the crystalline magnetic anisotropy $K_1$ we would estimate
$\chi_\perp(0)\approx 0.4$ 
(see Appendix \ref{app:single_domain}).
However, measurements by Valstyn, Hanton, \& Morrish (1962) 
of the permeability as a function of frequency
are better reproduced by a slightly larger value, $\chi_\perp(0)\approx0.6$,
which we adopt.
Our estimate for $\mu(\omega)$ for $\gamma$Fe$_2$O$_3$ is
shown in Fig.\ \ref{fig:Fe_3O_4_single}.
Valstyn et al.\ measured $\mu(\omega)$ for a powder of $0.03-0.2\micron$
diameter $\gamma$Fe$_2$O$_3$ spheres suspended in paraffin wax with a
volume filling factor 0.15; thus in effect there was a volume
filling factor of 0.10 for spheres with spontaneous magnetization
perpendicular to the applied oscillating field.
In Fig.\ \ref{fig:Fe_3O_4_single} we
show their experimental results for $\mu-1$
divided by 0.10.
The model which we are using appears to be in reasonably good agreement
with the measured frequency-dependent susceptibility for this material.

\subsection{Antiferromagnetic Materials \label{sec:antiferro}}

In antiferromagnetic materials
the exchange interaction leads to 
magnetic domains with magnetic moments which are ordered, 
but in such a way that
the net magnetization in the domain is zero.
The olivine fayalite (Fe$_2$SiO$_4$) is an example 
of an antiferromagnetic material.
These substances have zero frequency susceptibilities $\chi(0)$ 
which are similar to (but somewhat smaller than) those of normal
paramagnetic substances with similar concentrations of magnetic ions.
Our estimate for ``normal paramagnetism'' would therefore be a reasonable guide
to antiferromagnetic substances such as fayalite.

However, we do not expect interstellar grains to contain perfect
crystals; defects and impurities seem likely to be common, and the
materials may be amorphous (as appears to be the case for interstellar
silicates, as indicated by the profile of the interstellar 
$10\micron$ absorption feature due to the Si-O stretching mode).
Furthermore, if the magnetic material is in very small inclusions, 
perfect pairing of spins will not be possible along the boundary
of the inclusion.
For instance, Schuele \& Deetscreek (1962) find that 
small ($\ltsim10^{-6}\cm$) particles of NiO are
weakly {\it ferro-} or {\it ferri}magnetic although 
bulk NiO is {\it antiferro}magnetic.
Antiferromagnetism therefore seems unlikely -- we instead expect amorphous
fayalite, for example, to be at least weakly ferrimagnetic.
The high frequency response of such small inclusions would be determined
by the single-domain susceptibility (\S\ref{sec:Fe_incl}).

\section{Microwave Emission\label{sec:prediction}}

For a nonrotating 
grain whose {\it internal} degrees of freedom are in thermal equilibrium 
at a temperature $T$, the power radiated in frequency interval $d\nu$ 
is simply
\beq
P_\nu d\nu = 4\pi C_{abs}(\nu) B_\nu(T)
\eeq
where $C_{abs}(\nu)$ is the absorption cross section at frequency $\nu$,
and $B_\nu(T)$ is the Planck function.

We will use the term ``vibrational emission'' to refer to emission arising
from thermal fluctuations in the charge distribution, and therefore the
electric polarization, of the grain.
This will produce electric dipole emission.

We will use the term ``magnetic dipole emission'' to refer to emission
arising from thermal fluctuations in the magnetization of the grain material.
Estimation of this emission is one of the goals of the present paper.

We use the term ``rotational emission'' to refer to the emission arising
from rotation of the grain.
This is primarily due to the electric dipole moment which rotates with
the spinning grain (Draine \& Lazarian 1998a,b).
In principle, there will be rotational emission arising from the
rotating magnetic dipole moment of a spinning magnetized grain.
Only $a\ltsim10^{-7}\cm$ grains spin at frequencies $\nu\gtsim10\GHz$.
The magnetic moment of a spontaneously magnetized grain is
just $M_s V = (1/3)(4\pi M_s/10^3\G)(a/10^{-7}\cm)^3\debye$.
The largest value of $M_s$ is for metallic Fe 
(see Table \ref{tab:Feminerals}); an Fe grain would have a magnetic moment
$M_s V = 7.3 (a/10^{-7})^3\debye$, somewhat smaller than
the electric dipole moment $9.3 (a/10^{-7})^{3/2}\debye$ estimated
for neutral grains by Draine \& Lazarian (1998b).
Other magnetic materials have smaller values of $M_s$, and hence
the rotational emission at $\nu\gtsim10\GHz$ 
is expected to be dominated by electric dipole radiation.

\subsection{Candidate Materials\label{sec:known_materials}}

We consider 4 possible components of the interstellar grain population:
\begin{enumerate}
\item
$100\%$ of the Si and Fe incorporated
into amorphous silicate grains with paramagnetic behavior as in
Fig.\ \ref{fig:PM};

\item
$100\%$ of the Fe incorporated into small Fe$_3$O$_4$ grains;

\item 5\% of the Fe incorporated into bare Fe grains;

\item 5\% of the Fe incorporated into small metallic inclusions.

\end{enumerate}
For each case we assume a grain temperature $T=18\K$.
If $V$ is the volume per H atom for a grain component, then its
{\it magnetic dipole} contribution to 
the emissivity per H atom is 
\beq
{j_\nu\over n_\H} = {n_{gr}V\over n_H}
{4\pi h\nu^4\over c^3}{1\over [\exp(h\nu/kT)-1]}
{9\mu_2\over (\mu_1+2)^2 +\mu_2^2} ~~~,
\eeq
where we must remember that there may be additional electric dipole
emission [see eq.\ (\ref{eq:dipoleapprox}--\ref{eq:cabs_md})].

%\placefigure{fig:emissivity}
%{\begin{center}[Editor: Place Fig.\ \ref{fig:emissivity} here.]\end{center}}
\begin{figure}
\epsscale{0.90}
\plotone{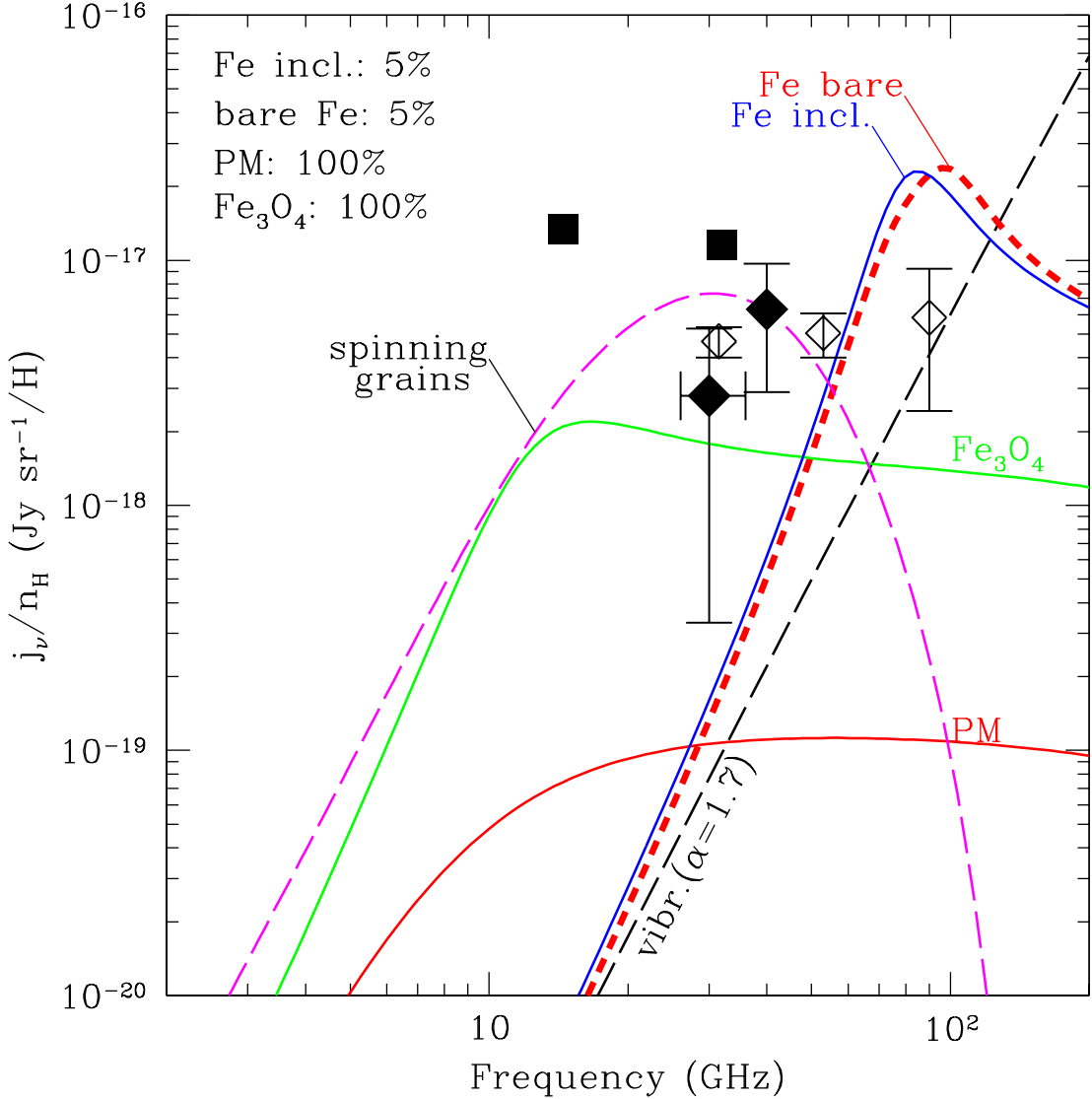}
\figcaption[f7.eps]{
	\label{fig:emissivity}
	Predicted thermal emission per H nucleon of interstellar dust
	for 
	(1) paramagnetic silicate
	grains containing 100\% of the Fe, Mg, and Si;
	(2) Fe$_3$O$_4$ grains containing 100\% of the Fe;
	(3) 5\% of the Fe in metallic iron, either as ``bare'' single-domain
	grains or as single-domain inclusions
	(see text).
	Also shown is the estimated ``vibrational'' electric dipole
	emission from interstellar dust with 
	$C_{abs}^{(ed)}\propto\nu^{1.7}$
	and the rotational emission from very small spinning grains
	as estimated for the ``Cold Neutral Medium''
	by Draine \& Lazarian (1998a,b).
	Observed emissivities are inferred from data of Kogut et al.\ (1996;
	open diamonds), de Oliveira-Costa et al.\ (1997; filled diamonds),
	and Leitch et al.\ (1997; filled squares).
	}
\end{figure}

The magnetic dipole emissivity for each grain component is shown in Figure 
\ref{fig:emissivity}; also shown is the estimate for the ``vibrational'' or
electric dipole emission, 

Sensitive studies of the microwave sky brightness have
revealed microwave emission from interstellar matter;
the emissivity per H
nucleon has been deduced from the cross-correlation of the microwave
sky brightness with
far-infrared emission from dust grains, using measurements from
the COBE DMR (Kogut et al.\ 1996),
the ground-based Saskatoon experiment (de Oliveira-Costa et al.\ 1997),
and OVRO (Leitch et al.\ 1997).
These observational results are shown in Fig.\ \ref{fig:emissivity}.

We see that a population of Fe grains or grains with Fe inclusions
would be predicted to produce very strong emission near $70\GHz$,
with the peak occurring near the Fr\"ohlich resonance for Fe spheres.
Even if the permeability of Fe has been estimated correctly, this
peak would be broadened if, as expected, the Fe grains or inclusions
are not all spherical, since the Fr\"ohlich resonance frequency
depends upon the particle shape.
The $90\GHz$ emission reported by Kogut et al.\ (1996) appears to
limit the amount of interstellar Fe in metallic form to perhaps
$\ltsim 5\%$ of the total Fe.
In particular, this limits the fraction of interstellar Fe which
can be present in Fe-Ni inclusions such as found in ``GEMS''
(Bradley 1994; Martin 1995).

\subsection{Hypothetical Materials\label{sec:hypothetical_materials}}

We have estimated the emission expected from pure iron and magnetite,
but the true state of Fe in interstellar grains is not known.
It is possible that the Fe is concentrated in a form which is 
magnetic -- not so strongly as pure iron, but more strongly
than ferrimagnetic magnetite.
This could, for example, be an Fe/Ni alloy with an appreciable concentration
of O, H, Si, or other impurities.
We now ask whether it is {\it possible}
for magnetic dipole emission to account for the observed emission in
the 14-90 GHz range.
We approach this question by seeking to
``customize'' the magnetic properties, within the range of reasonable
parameters, to see whether there are possible values which would lead
to the observed emission.
We make the following assumptions:
\begin{enumerate}
\item The magnetic material is in (spontaneously-magnetized)
	single-domain grains, with a susceptibility $\chi_\perp$ 
	perpendicular to the
	direction of spontaneous magnetization.
\item The frequency dependence of the susceptibility is given by 
	eq.\ (\ref{eq:chi_cd}).
\item The low frequency susceptibility $\chi_\perp(0)$ should be 
	somewhat smaller than the value $\chi_\perp(0)=3.3$ which
	we have estimated for pure iron.
\item The spontaneous magnetization should be appreciably smaller than
	the value $4\pi M_s=22$kG for pure iron.  As a result, the
	characteristic gyrofrequency $\omega_0/2\pi = (e/m_e c)(2 M_s/3)$ 
	should be appreciably
	smaller than the $\sim20\GHz$ value estimated
	for Fe.
\item We assume that nearly 100\% of the interstellar Fe is in the
	hypothetical material $X$, and
	assume the Fe within the material $X$ to contribute a mass
	density of $4\g\cm^{-3}$ 
	[thus the volume per H atom of the hypothetical
	material is $V_X = 7.5\times10^{-28}\cm^3/\H$ --
	see eq.\ (\ref{eq:V_Y})].
\end{enumerate}

\begin{figure}
\epsscale{0.90}
\plotone{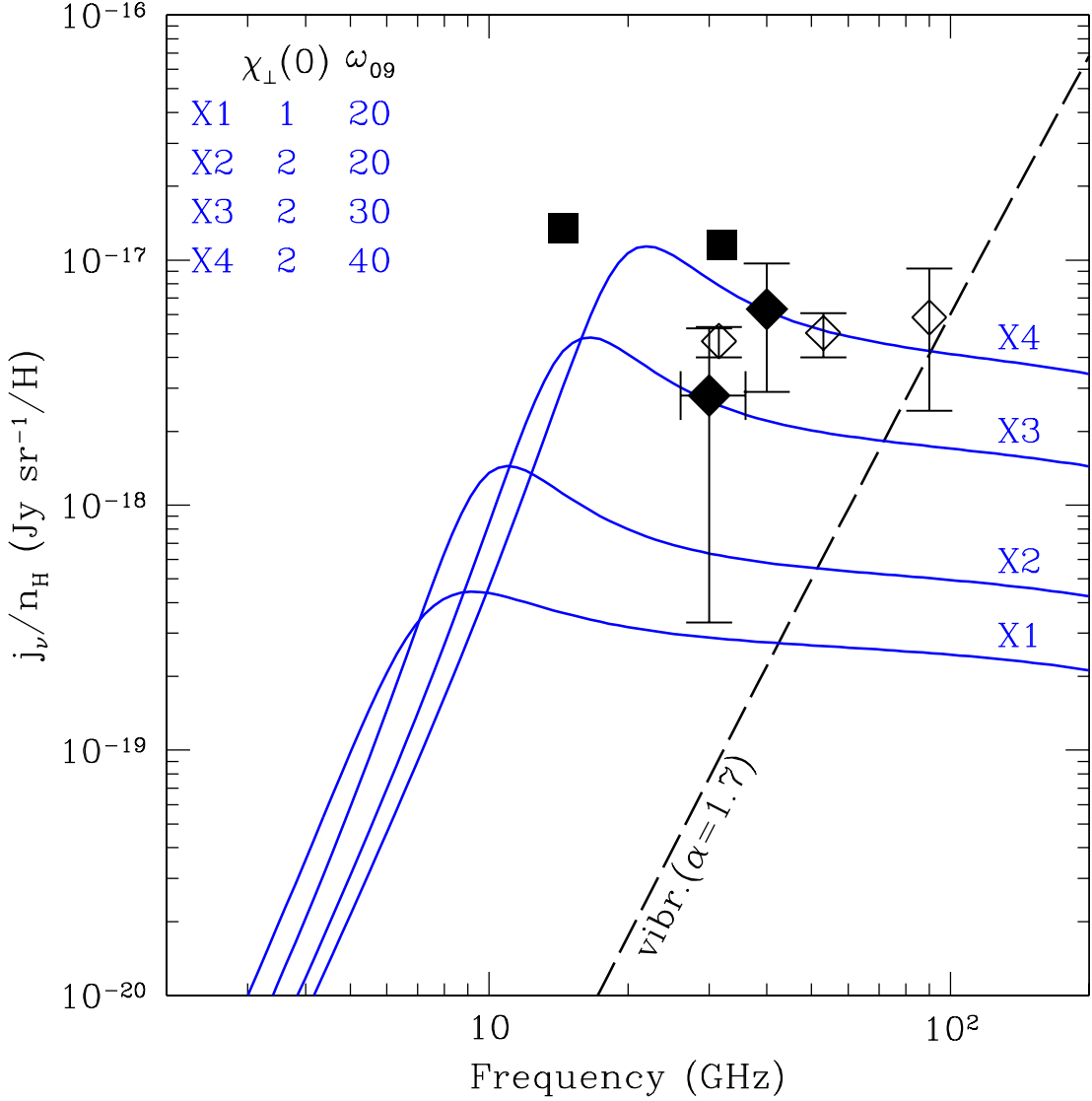}
\figcaption[f8.eps]{
	\label{fig:custom_emissivity}
	Predicted thermal emission per H nucleon of interstellar dust
	for hypothetical Fe-rich materials X1-X4.  In each case we
	assume that a fraction $f_X=1$ 
	of interstellar Fe is in a substance
	with $\rho_Xz_x=4\g\cm^{-3}$ 
	[see eq.\ (\protect{\ref{eq:V_Y}})],
	with an assumed static susceptibility $\chi_0$ and
	characteristic gyrofrequency $\omega_0=10^9\omega_{09}\s^{-1}$.
	Observed emission is as described in 
	Fig.\ \protect{\ref{fig:emissivity}}.
	Thermal emission from hypothetical material X4 could
	account for a substantial fraction of
	the observed 14-90 GHz emission.
	Also shown is the estimated ``vibrational'' electric dipole
	emission from interstellar dust with 
	$C_{abs}^{(ed)}\propto\nu^{1.7}$.
	}
\end{figure}

Calculations are shown for 4 hypothetical materials, denoted X1, X2, X3, and
X4, with $\chi_\perp(0)$ and $\omega_0$ values as shown in 
Fig.\ \ref{fig:custom_emissivity}.
In Fig.\ \ref{fig:custom_emissivity} we see that in order to contribute
an appreciable fraction of the observed 14-90 GHz emission, the hypothetical
magnetic material must have $\chi_\perp(0)\approx 2$ and
$\omega_0/2\pi\approx 6\GHz$.
Note that material X1 does not have a Fr\"ohlich resonance, and 
materials X2,X3,X4, with $\chi_\perp(0)=2$, have only a weak
Fr\"ohlich resonance (see footnote \ref{fn:Frohlich}).

%\placefigure{fig:custom_emissivity}
%{\begin{center}[Editor: Place Fig.\ \ref{fig:custom_emissivity} here.]\end{center}}

We conclude that {\it if} a large fraction of the interstellar Fe were
incorporated into a moderately strong magnetic material with properties
approximating those of our hypothetical material X4, then the thermal
emission from these grains would approximately reproduce the observed
thermal emission in the 14-90 GHz range.
We stress that we are {\it not} arguing that such material exists -- indeed,
we have previously shown (Draine \& Lazarian 1998a,b) 
that the observed emission seems likely to be due
to rotational emission from a population of ultrasmall grains believed
to be present for independent reasons.
The point of the present discussion is to show that one cannot exclude the
possibility that magnetic grains make an appreciable contribution to the
emission in this frequency range.
This issue should be resolved through observations of dark clouds
(see \S\ref{sec:discussion}).

\section{Polarization\label{sec:polarization}}

From Fig.\ \ref{fig:emissivity}
it is seen that
if a small fraction of interstellar Fe is in iron grains, these could
account for a substantial fraction of the diffuse emission in the
$50-100\GHz$ region.
Alternatively, we see in Fig.\ \ref{fig:custom_emissivity}
that
if a large fraction of the Fe is in a magnetic
material with the properties of our hypothetical material ``X4'',
then a substantial fraction of the 14-90 GHz emission could
be magnetic dipole emission from such grains.

In recent years there has been significant progress toward understanding
the alignment of interstellar grains.
Analysis of the ``crossover'' phenomenon for
grains subject to superthermal rotation now indicates that even
ordinary paramagnetic relaxation may suffice to align larger
($a\gtsim10^{-5}\cm$) grains (Lazarian \& Draine 1997), but
radiative torques due to starlight appear to dominate those due to
paramagnetic relaxation (Draine \& Weingartner 1997).
If smaller ($a\ltsim10^{-5}\cm$) strongly magnetic grains are present
(e.g., iron or our hypothetical material X4), it appears
likely that ferromagnetic relaxation would effectively align their angular
momenta with the galactic magnetic field $\bB_0$, 
so that their ``long'' axes would tend to be perpendicular to $\bB_0$.
If such grains are present, we then expect frequency-dependent
polarization of the microwave emission from interstellar dust.

To estimate the likely polarization, we assume the grains to
be ellipsoids with semiaxes
$a_1\leq a_2\leq a_3$, spinning with the short axis $\hat{a}_1$
parallel to the angular momentum $\bJ$
(as expected for suprathermally rotating grains -- see Purcell 1979).
We assume ``perfect'' alignment of $\bJ$ with the galactic magnetic field
$\bB_0$, with $\bB_0$
perpendicular to the line-of-sight;
the polarization of emitted radiation will be taken to be positive when
the electric vector is perpendicular to $\bB_0$.
For imperfect alignment, the polarization for perfect alignment
should be multiplied by the ``Rayleigh reduction factor'' (Lee \& Draine 1985)
\beq
R=(3/2)(\langle\cos^2\theta\rangle-1/3) ~~~,
\eeq
where $\theta$ is the angle
between $\bB_0$ and $\bJ$.

The predicted polarization depends on assumptions concerning the grain
structure.
We will consider two limiting cases.

\subsection{Single Domain Grains\label{sec:pol_single_domain}}

Suppose the grains which dominate the $10-100\GHz$ emision 
each consist of a single
magnetic domain.\footnote{
	Recall that Fe grains smaller than $a_c\approx3\times10^{-6}\cm$ 
	always consist of a single domain (Morrish 1980).
	}
We will assume the grains to have spontaneously magnetized with ${\bf M}_s$
along the long axis $\hat{a}_3$
(this minimizes the magnetic energy).
As discussed above in \S\ref{sec:Fe_single}, the permeability $\mu$ is
anisotropic: $\mu=1$ for $\bH\parallel\bM_s$,
and $\mu=\mu_\perp$ for $\bH\perp\bM_s$.
Thus the degree of polarization $P=(I_e-I_h)/(I_e+I_h)$ where
\beq  
I_e=
{1\over 2}\left[ 
{\epsilon_2  \over |1+L_2(\epsilon-1)|^2}
+
{\epsilon_2  \over |1+L_3(\epsilon-1)|^2}
\right]
+
{\mu_{\perp,2}       \over |1+L_1(\mu_\perp-1)|^2}
\eeq
\beq
I_h=
{\epsilon_2 \over |1+L_1(\epsilon-1)|^2}
+
{1\over2}{\mu_{\perp,2} \over |1+L_2(\mu_\perp-1)|^2} ~~~,
\label{eq:I_e2}
\eeq
The dielectric function of metallic iron can be approximated at
low frequencies by a Drude model with $\omega_0=0$:
\beq
\epsilon(\omega)\approx {i(\omega_p\tau)^2 \over \omega\tau-i(\omega\tau)^2}
\eeq
where $\tau\approx 3.8\times10^{-14}\s$ and $\omega_p\tau\approx200$.
This approximately reproduces the tabulated values of $\epsilon$
(Palik 1991).  With this dielectric constant, the microwave emission
is dominated by magnetic dipole radiation 
($|\epsilon|>10^5$ for $\nu<100\GHz$), 
so that $C_{abs}^{(ed)}\ll C_{abs}^{(md)}$
and the precise value of $\epsilon$ is not critical.

\begin{figure}
\epsscale{0.90}
\plotone{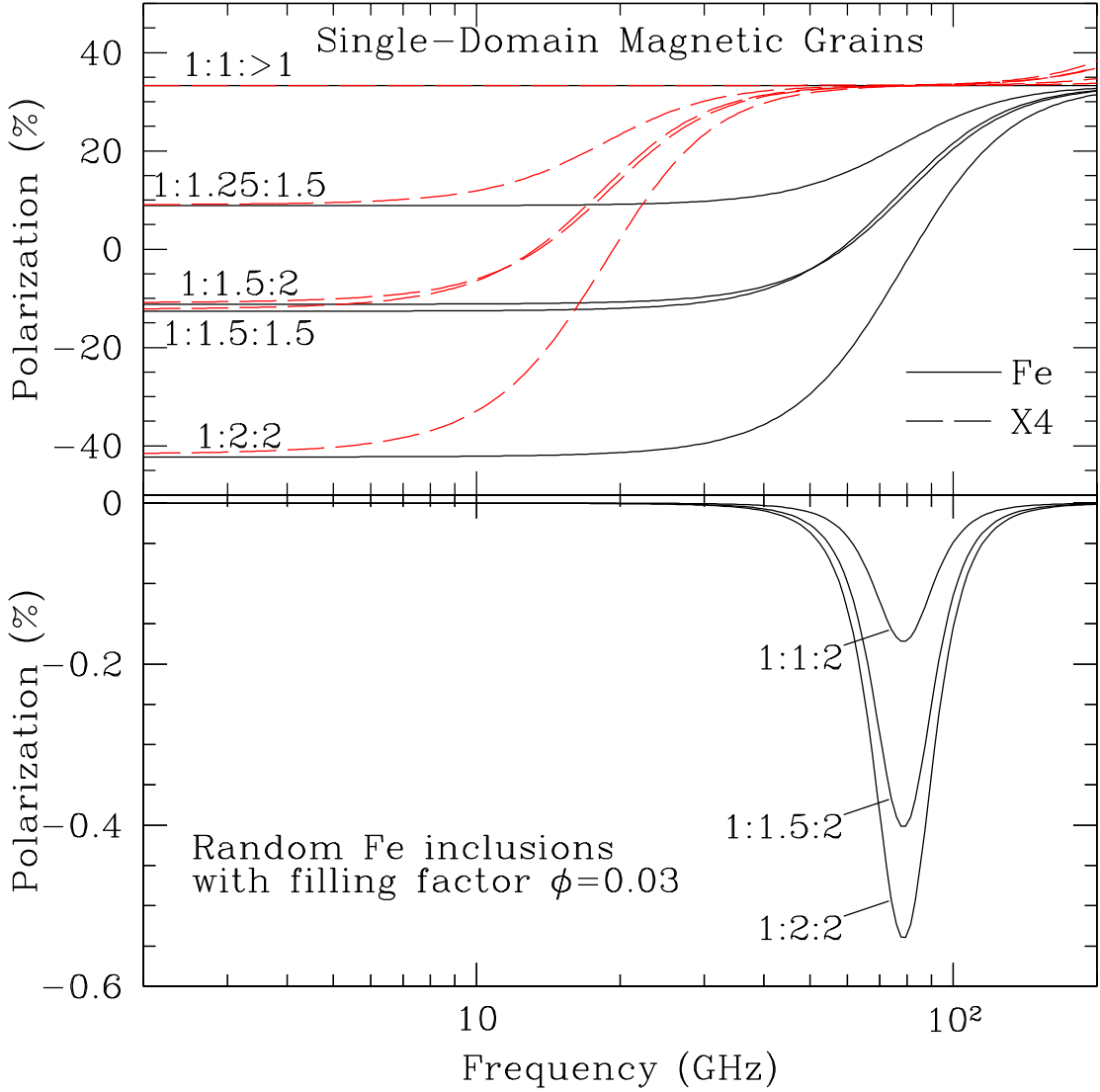}
\figcaption[f9.eps]{
	\label{fig:polarization}
	Polarization of thermal emission from single-domain grains
	(upper panel)
	consisting of either metallic Fe (solid lines) 
	or the hypothetical material X4 (broken lines),
	assuming perfect
	(spinning) alignment (see text),
	Single-domain prolate spheroids 
	($a_1\!:\!a_2\!:\!a_3\!::\!1\!:\!1\!:\!>1$) 
	have $P=1/3$ for $\nu\ltsim 100\GHz$.
	The lower panel shows the polarization of the thermal emission
	from perfectly aligned ellipsoids composed of material with
	metallic Fe inclusions with
	volume filling factor $\phi=0.02$.
	Results are shown for various values of the axial ratios
	$a_1\!:\!a_2\!:\!a_3$.
	}
\end{figure}

The resulting frequency-dependent polarization $P$ is shown in 
Fig.\ \ref{fig:polarization} for different grain shapes.
For both Fe or X4 grains, 
at high frequencies ($\nu\gtsim100\GHz$ for Fe,
$\nu\gtsim 25 \GHz$ for X4) 
the polarization is large ($P=1/3$) and
positive (i.e., $\bE$ perpendicular to the ``short axis'' of the grain).
At these high frequencies $|\mu_\perp-1|^2\ll1$ so that the 
values of the ``shape factors''
$L_1>L_2>L_3$ are unimportant --
the calculated polarization arises because the thermal fluctuations in the
magnetization are perpendicular to $\hat{a}_3$; since $\hat{a}_3\perp\bJ$
the magnetic dipole emission tends to have $\bH\parallel\bJ$, and hence
$\bE\perp\bJ$, for ``positive'' polarization.

%\placefigure{fig:polarization}
%{\begin{center}[Editor: Place Fig.\ \ref{fig:polarization} here.]\end{center}}

At lower frequencies ($\ltsim50\GHz$ for Fe,
$\ltsim 14\GHz$ for X4) $|\mu-1|>1$ and shape effects now
matter.
There are still no magnetic fluctuations parallel to $\hat{a}_3$,
but if $a_1$ is sufficiently small compared to $a_2$, $L_1$ will
be considerably larger than $L_2$, and the magnetic dipole emission will
tend to have $\bH\perp\hat{a}_1$, hence $\bE\parallel\hat{a}_1$, for
``negative'' polarization.
This explains the behavior seen in Fig.\ \ref{fig:polarization}.

\subsection{Grains With Magnetic Inclusions}

Suppose the grains to be ellipsoids containing isotropically-oriented
magnetic inclusions with filling factor $\phi\approx0.03$, as found
for GEMS (see \S\ref{sec:mag_grains}).
Since the grain material is now isotropic, the polarization is given
by eq.\ (\ref{eq:pol}-\ref{eq:I_h}).

If the inclusions are metallic Fe, and contain $N>10^4$ atoms, we can
use the permeability $\mu(\omega)$ shown in Fig.\ \ref{fig:SPM}.
The resulting polarization for perfectly-aligned spinning grains is
shown in Fig.\ \ref{fig:polarization}.
Thermal emission from these grains is predominantly magnetic dipole
radiation; since the grain material is isotropic\footnote{If
	the number of magnetic inclusions in a grain is small,
	the material within each grain will not, strictly speaking,
	be isotropic.  
	However, if the orientations of individual inclusions are
	statistically independent of the overall grain shape, it
	is appropriate to approximate the grain material as isotropic
	in order to estimate the emission from the grain ensemble.
	}
	
the emitted radiation
tends to have $\bH$ perpendicular to the short axis $\hat{a}_1$.
However, since $|\mu-1|\ll 1$
(see Fig.\ \ref{fig:SPM}), the $L_j$-dependent term in 
eq.\ (\ref{eq:I_e}-\ref{eq:I_h}) is only a small correction; hence
$I_e\approx I_h\approx (\epsilon_2+\mu_2)$ and the
polarization $P=(I_e-I_h)/(I_e+I_h)$ is small.

\section{Discussion\label{sec:discussion}}

Fe is the 9th most abundant element by number 
(after H, He, O, C, N, Ne, Mg, and Si),
and 5th by mass (after H, He, O, C),
and nearly all of the interstellar Fe is in dust grains.
It is therefore inevitable that some fraction of the interstellar grain
material must be quite Fe-rich.
We don't know what mineral form the Fe mainly resides 
in (a number of possibilities
are given Table \ref{tab:Feminerals}), but the material
will be at least paramagnetic, and quite possibly 
ferrimagnetic or ferromagnetic.

It is therefore important to consider the emission of electromagnetic
radiation as a result of thermal fluctuations in the magnetization of
the grain material.
These fluctuations take place at microwave frequencies and below (the 
magnetization is constant on timescales short compared to the time for
electrons to precess in the internal magnetic fields within the material;
these precession frequencies are at most $\sim 20\GHz$ (the value for
metallic iron).
As a result, {\it thermal} magnetic dipole emission from interstellar 
grains 
$\sim 100\GHz$ may be be stronger than the thermal electric
dipole ``vibrational'' emission
(arising from thermal fluctuation is in the charge distribution within the
grain) at frequencies $\ltsim 100\GHz$.\footnote{Rotational
	electric dipole radiation from rapidly-rotating ultrasmall
	grains is also important at these frequencies 
	(Draine \& Lazarian 1998a, 1998b).
	}
We show how this magnetic dipole emission can be calculated, but the
emission requires knowledge of the magnetic permeability at these
frequencies.  We have estimated this permeability for various
materials of interest, including metallic iron and magnetite (Fe$_3$O$_4$),
but have been required to extrapolate to the high frequencies of 
interest.
It would be of great value to have direct measurements of
the magnetic properties of small Fe particles at 50-100 GHz.

According to our estimates, metallic iron would produce such strong emission
near $\sim 70\GHz$ that not more than $\sim5\%$ of the Fe can be in the
form of pure metallic iron in order not to exceed the observed emission
at these frequencies.
This limits the fraction of interstellar grains with Fe-Ni inclusions
such as in the GEMS found by Bradley (1994).

Magnetite, on the other hand, would radiate only a fraction of the power
emitted by interstellar grains in the 14-90 GHz region.
However, it is possible that interstellar Fe could be present in some
Fe-rich substance (e.g., an Fe/Ni alloy with an appreciable concentration
of  Mg,Si,O,H impurities) 
with magnetic properties intermediate between those of metallic iron and
magnetite, and we therefore cannot rule out the possibility that the bulk
of the observed emission in the 14-90 GHz range could be thermal
magnetic dipole radiation.
It would be of great value to have laboratory measurements on various
plausible candidate materials (e.g., amorphous olivine, or Fe-containing
alloys) at these
frequencies.

Rotational emission from spinning dust grains has previously been
proposed as the explanation for the observed 15-90 GHz emission from
dust in diffuse clouds (DL98a,b).
There are two ways in which this new emission mechanism -- magnetic
dipole radiation from thermal fluctuations in grain magnetization -- can
be distinguished from the hypothesized rotational emission:
\begin{itemize}
\item The rotational emission requires ultrasmall grains.  Since we
	have reason to believe that ultrasmall grains are depleted in
	dense regions, we would then expect the rotational emission
	to be relatively weak in dense gas.
	On the contrary, thermal magnetic dipole emission from dust
	should be largely unaffected by coagulation of the dust grains.
	Therefore, observations of (or upper limits on) the $10-60\GHz$
	emission from dense clouds can be distinguish between rotational
	emission from ultrasmall grains or thermal magnetic dipole 
	emisson from magnetic grain materials.

\item If produced largely by single-domain grains, 
	the magnetic dipole emission potentially has a complex and strong
	polarization signature, which could be very different from
	the polarization expected from
	spinning dust grains which have been partially-aligned
	by magnetic dissipation (Lazarian \& Draine 1998).
\end{itemize}

\section{Summary\label{sec:summary}}

The principal results of this paper are as follows:

\begin{enumerate}
\item Formulae are presented for electric and magnetic dipole
absorption cross sections for homogeneous spheres or ellipsoids 
of material with
dielectric function $\epsilon$ and magnetic permeability $\mu$.

\item The Kramers-Kronig relation relating the total grain volume
to $\int_0^\infty C_{ext}d\lambda$ is generalized to include the case
of magnetic grains.
The resulting eq.\ (\ref{eq:KKresult}) shows that if an appreciable
fraction of the grain material is magnetic, then magnetic dipole
effects must dominate the extinction cross section at the
frequencies $\nu\ltsim 30\GHz$ where plausible grain materials have
a magnetic response.

\item Because most of the Fe is in solid form in the interstellar medium,
it is expected that some fraction of the interstellar grain population
must be appreciably magnetic -- either paramagnetic, superparamagnetic,
ferrimagnetic, or ferromagnetic.
The frequency-dependent magnetic properties of candidate materials 
are estimated.

\item The magnetic dipole emission from either 
paramagnetic grains or magnetite (Fe$_3$O$_4$) grains is well
below the expected electric dipole emission due to vibrational modes
at $\nu\gtsim100\GHz$, or rotational emission from very small
grains at $\nu\ltsim100\GHz$.

\item Fe grains or inclusions will have the magnetic analogue of
a Fr\"ohlich resonance at $\sim70\GHz$, where the magnetic
dipole absorption cross section peaks because $\mu_1=-2$.
This will result in strong magnetic dipole absorption from
ferromagnetic iron particles near $\sim70\GHz$.

\item If our estimate for the magnetic susceptibility of single-domain
Fe is correct, not more than $\sim5\%$ of interstellar Fe can be
in the form of metallic iron; otherwise the thermal magnetic dipole
emission from interstellar dust at $90\GHz$ would greatly exceed
the emission from dust measured by Kogut et al.\ (1996).

\item If a substantial fraction of interstellar Fe is in a moderately
strong magnetic material with the properties resembling our
hypothetical material ``X4'' in Fig.\ \ref{fig:custom_emissivity},
then thermal emission from such grains could account for a substantial
fraction of the observed 14-90 GHz emission.
At this time this possibility cannot be excluded, even though
the 14-90 GHz emission seems most likely due to spinning ultrasmall
dust grains (Draine \& Lazarian 1998a,b).

\item If nonspherical single-domain 
ferromagnetic grains are present, the
magnetic dipole emission will be polarized.
Such grains are expected to be spontaneously magnetized along the
``long'' axis.
The polarization is expected to depend strongly on frequency,
with ``normal'' polarization (electric vector perpendicular to $\bB_0$)
at frequencies $\gtsim 100\GHz$, but with a decrease, and perhaps
even reversal (depending upon grain shape), 
of the polarization for $\ltsim 30\GHz$.

\end{enumerate}

\acknowledgements

We thank
Phil Myers
for helpful discussions,
and
Robert Lupton for the availability of the SM package.
B.T.D. acknowledges the support
of NSF grant AST-9619429, and
A.L. the support of NASA grants NAG5-2858 and NAG5-7030.

\appendix
\section{Microwave Absorption due to Stark-Effect Splitting
	\label{app:stark_effect}}

The electric fields within a solid -- whether crystalline or
amorphous -- can split the magnetic sub-levels of paramagnetic ions 
in the solid.
Kittel \& Luttinger (1948) noted that while these splittings 
usually correspond to frequencies $\sim 10^{13}\Hz$,
under some circumstances the splittings
correspond to microwave frequencies.
To estimate the magnitude of the associated absorption,
suppose that the solid contain a density $dn$ of species with
energy levels split by frequencies in the interval $d\omega$, and
let $|\mu_{ul}|^2$ be the magnetic dipole matrix element between
the upper and lower states,
giving a spontaneous decay rate 
\beq
A_{ul}={4\omega^3\over3\hbar c^3}|\mu_{ul}|^2
~~~.
\eeq
If $\omega \ll kT/\hbar$, then $d(n_l-n_u)=(\hbar\omega/2kT)dn$.
The absorption coefficient is
\beq
\alpha = 
{2\pi^2\over3}\left({\hbar\omega\over kT}\right){dn\over d\ln\omega}
{|\mu_{ul}|^2\over \hbar c} ~~~,
\eeq
corresponding to an imaginary component of the magnetic susceptibility
\begin{eqnarray}
\mu_2 &=& {c\over\omega}\alpha = 
{2\pi^2\over3}\left({|\mu_{ul}|^2\over kT}\right)
{dn\over d\ln\omega}\\
&=& 2.7\times10^{-3} \left({|\mu_{ul}|^2\over \mu_B^2}\right)
\left({15\K\over T}\right)
\left({dn/d\ln\omega\over10^{22}\cm^{-3}}\right) ~~~.
\label{eq:mu_stark}
\end{eqnarray}
where $\mu_B$ is the Bohr magneton.
The total atomic density $\sim10^{23}\cm^{-3}$ in the solid,
so $dn/d\ln\omega=10^{22}\cm^{-3}$ would correspond to 10\% of
the atoms having resonances within a frequency interval
$\Delta\ln\omega=1$.
Such a high value would probably only occur if the Stark-effect
splittings are concentrated near a few frequencies.

\section{Susceptibility of Single-Domain Ferromagnetic Particles
	\label{app:single_domain}
	}
Consider a spherical particle consisting of a single domain.
At temperatures $T \ltsim T_C/2$, where $T_C$ is the Curie
temperature, we may take the spontaneous magnetization to be approximately
equal to the saturation magnetization $M_s$.
Suppose the crystal has cubic symmetry.
The free energy is a function of the direction of the magnetization.
If $\alpha_1$, $\alpha_2$, and $\alpha_3$ are the direction cosines
of the magnetization relative to the crystal axes,
the ``anisotropy'' free energy is
\beq
F_K=K_1(\alpha_1^2\alpha_2^2+\alpha_2^2\alpha_3^2+\alpha_3^2\alpha_1^2)
+ K_2\alpha_1^2\alpha_2^2\alpha_3^2
\eeq
where $K_1$ and $K_2$ are the anisotropy constants (Morrish 1980).

\subsection{$K_1>0$, $K_2 >0$}

If $K_1>0$ and $K_2>0$, then the ``easy'' directions are along the
cubic axes (e.g., $\langle 100\rangle$).
If we now apply a weak magnetic field $H_\perp$ 
perpendicular to the magnetization,
the magnetization will deflect by an angle $\theta$.
For weak fields, $\theta\ll 1$ and
the anisotropy energy and magnetic energy become
\beq
F_K = K_1 \theta^2 + O(\theta^4)
\eeq
\beq
F_H = -{\bf M}\cdot{\bf H}=
M_sH_\perp\theta + O(\theta^2) ~~~.
\label{eq:F_H}
\eeq
Minimizing $(F_K+F_H)$, we find 
$\theta=M_s H_\perp / 2 K_1$.
Thus, the transverse susceptibility $\chi_\perp=M_s\theta/H_\perp$ is
\beq
\chi_\perp (0) \approx {M_s \theta \over H_\perp} = 
{M_s^2 \over 2 K_1} ~~~.
\label{eq:chi_perp_100}
\eeq
For Fe, $M_s=1750 {\rm\, G}$, and $K_1=4.6\times10^5\erg \cm^{-3}$
(Morrish 1980), thus
$\chi_\perp(0)=3.3$ and
$\mu_\perp(0)=1+4\pi\chi_\perp(0)=43$.
This is significantly smaller than the value $\mu(0)\approx 150$ 
characterizing
bulk (multidomain) Fe.

\subsection{$K_1<0$, $K_2<0$}

When $K_1<0$, $K_2<0$, the ``easy'' directions are along the diagonals
(e.g., $\langle 111 \rangle$).
The anisotropy energy is now
\beq
F_K= K_1 \left( {2\over 9} - {2\over 3}\theta^2 \right)
+ K_2\left( {1\over 27} - {2\over 9}\theta^2 \right) + O(\theta^4) ~~~,
\eeq
with $F_H$ again still by eq.(\ref{eq:F_H}).
Again minimizing $(F_K+F_H)$ we find
$\theta=-9 M_s H_\perp/[12K_1+4K_2]$, and
\beq
\chi_\perp(0) \approx {-9M_s^2 \over 12 K_1 + 4 K_2}
\label{eq:chi_perp_111}
\eeq

Maghemite ($\gamma$Fe$_2$O$_3$) is cubic with
$4\pi M_s=4780\G$ (Dunlop \& \"Ozdemir 1997).
The magnetic properties do not appear to have been measured at low
temperatures; at room temperature,
$K_1=-2.5\times10^5\erg\cm^{-3}$ (Valstyn, Hanton, \& Morrish 1962),
from which we estimate
$\chi_\perp(0) \approx 0.43$.
As discussed in \S\ref{sec:ferri} above, a slightly larger value
of $\chi_\perp(0)\approx0.6$ is in agreement with measurements
by Valstyn et al.\ (1962)
at GHz frequencies.

\subsection{Oscillating Fields}

The above discussion applies to static fields.  For oscillating fields,
we estimate $\chi_\perp(\omega)$ using eq. (\ref{eq:chi_cd}),
with $\omega_0=(e/m_e c)(4\pi M_s/3)$, the precession frequency of
an electron in the internal field $4\pi M_s/3$.

\end{document}